\documentclass[%
reprint,
amsmath,amssymb,
aps
]{revtex4-2}

\usepackage{amsmath}%
\usepackage{amssymb}
\usepackage{graphicx}
\usepackage{dcolumn}
\usepackage{bm}
\usepackage{color}

\newcommand{\gcc}{\mbox{g~cm$^{-3}$}}
\newcommand{\xg}{x_{g}}

\newcommand{\B}{\bm{B}}

\newcommand{\vu}{\bm{u}}

\newcommand{\msun}{M_\odot}
\def\dy{\textcolor{black}}
\def\dgy{\textcolor{black}}

\begin{document}

\title[Zeeman effect in oscillations of magnetars]{Zeeman effect in oscillations of magnetars with toroidal magnetic fields}

\author{D. G. Yakovlev}
\email{yak.astro@mail.ioffe.ru}
\affiliation{Ioffe Institute, Politekhnicheskaya street 26, Saint Petersburg, 194021, Russia}%
\author{I. E. Fedorov}%
\affiliation{Murino Educational Center Nr.\ 2, Mendeleev boulevard 20/1, Murino, 
Leningrad region, 188662, Russia}%

\date{\today}

\begin{abstract}
Magnetars are neutron stars with superstrong magnetic fields. Some of them (soft-gamma repeaters, SGRs) demonstrate gigantic flares which nature is still
unclear. At decay phase of such flares one often observes quasi-periodic
oscillations (QPOs) which are treated as stellar oscillations triggered by the flares.
We study, for the first time, magneto-elastic oscillations
of magnetars possessing toroidal magnetic fields confined in the
stellar crust, without imposing axial symmetry of perturbations. We show that the Zeeman effect makes the oscillation spectrum much richer than for axially symmetric oscillations. 
The main properties of theoretical
QPO spectra are discussed as well as their potential to interpret observations
and explore magnetar physics.   
\begin{description}
\item[PACS numbers]
97.60.Jd, 
04.40.Dg, 
98.70.Rz, 
32.60.+i 
\end{description}
\end{abstract}

\maketitle

\section{Introduction}
\label{s:introduct}

Zeeman splitting of quantum energy levels by a magnetic field
is a famous physical effect. It is remarkable that it can be
observed on macrophysical level in such exotic objects as magnetars
which are neutron stars with superstrong 
magnetic fields. The Zeeman effect can split oscillation frequencies of magnetars which is useful for exploring basic
properties of these objects.

Neutron stars attract great attention because they contain matter
under extreme physical conditions; many properties of this matter are
still not clear (e.g. Refs. \cite{ST1983,HPY2007}). Generally, neutron stars consist of two main layers: the outer layer often called the crust, 
and the core. For a typical neutron star \dy{with the mass $M \sim 1.4\,\msun$ ($\msun$ being
the solar mass) the} radius is $R \sim 12$ km. The crustal matter consist mainly
of ions (atomic nuclei), electrons and (at densities $\rho \gtrsim 4 \times 10^{11}$ $\gcc$) free neutrons.  The crust depth is $\sim 1$ km, and its mass is $\sim 0.01\, \msun$. At the crust bottom the density reaches about one half of the saturation density of nuclear matter (the latter being about 
$\rho_0 \approx 2.8 \times 10^{14}$ 
$\gcc$). As a rule, atomic nuclei in the crust form Coulomb crystal.
Under the crust, there is a bulky and massive core containing superdense
nuclear matter. The central density reaches several $\rho_0$.

Magnetars form a special class of neutron stars (see Ref.\ \cite{2017KB} for a comprehensive review). Many problems of their structure and evolution are still under debates employing rather  controversial models, although it is generally believed that the presence of strong magnetic fields is crucial. We will be interested in an important group of magnetars called
soft-gamma repeaters (SGRs). Occasionally they demonstrate extremely powerful flares (with total energy release up to
$\sim 10^{46}$ erg). At the decay phase of these flares one often detects quasi-periodic oscillations (QPOs) of X-ray and
soft gamma-ray emission with typical frequencies from a few tens Hz to several kHz. 

Neither the mechanism of triggering the flares nor the mechanism of QPO generation are known, and we will not study them here. Nevertheless, it is commonly thought that the flares are regulated by strong magnetic fields (e.g. Ref.\ \cite{2017KB}), and the QPOs are associated with oscillations of magnetars excited in the flares. 

The QPOs were theoretically predicted in Ref.\
\cite{1998Duncan}. They were discovered \cite{2005Israel,2005Strohmayer,2006Watts} in 2005--2006 
after processing the observations
of the giant flare of SGR 1900+14 
(27/09/1998) and the hyperflare of SGR 1806--20 (27/12/2004). These observations, as well later observations of other
SGRs, have been processed and reprocessed 
(e.g., \cite{2011Hambaryan,2014Huppen,2014Huppenkothen,2018Pumpe}). 

\dy{Magnetar QPOs are usually treated as 
torsional neutron star pulsations  associated with sufficiently slow 
elastic shear perturbations in crystallized crust and/or Alfv\'en 
perturbations due to elasticity of magnetic field lines. Their typical propagation speeds are, respectively, the shear and Alfv\'en velocities. Shear perturbations exist in the crystallized crust, but Alfv\'en perturbations are not necessarily confined there: they can propagate to other 
magnetized regions of stars. 
In the absence of magnetic field, they are purely torsional shear modes. In the presence of the field, they are called torsional magneto-elastic, or just magneto-elastic modes}. In very high fields crystal elasticity may
become unimportant.

The theory of torsion \dy{shear} oscillations was began in 1980s
\cite{1980Hansen,1983ST,1988McDermott}. It was 
further elaborated after the
discovery of QPOs; see, e.g. Refs.\ 
\cite{2007Samuel,
2009Andersson,
2012Sotani,
2013aSotani,
2013Sotani,
2016Sotani,
2017aSotani,
2017Sotani,
2018Sotani,
2019Sotani,
2020KY,
2023Yak2},
simultaneously with rapidly developing theory of magneto-elastic oscillations; e.g. Refs.\
\cite{2006Levin,
2006Glampeda,
2007Sotani,
2007Levin,
2008Sotani,
2008Lee,
2009Colaiuda,
2009CD,
2011Gabler,
2011vanHoven,
2011Colaiuda,
2012vanHoven,
2012Gabler,
2012Colaiuda,	
2013Gabler,
2013Gabler1,
2014Passamon,
2016Link,2016Gabler,2018Gabler,2024Chamel}. 
The latter publications have been mostly restricted to
studying axially symmetric magnetic fields and deformations of fluid elements. 
These studies neglected Zeeman splitting of oscillation
frequencies by magnetic fields which made the family of 
theoretical frequencies incomplete.

The Zeeman splitting of magnetar QPOs was first discussed and properly estimated by Shaisultanov and Eichler \cite{2009SE} in
2009, but their publication did not
attract much attention.
Two other 
publications \cite{2023Yak1,2024Yak} on the subject, devoted to 
magnetars with dipole magnetic fields in the crust, appeared
only recently. They demonstrated, using the first-order
perturbation theory with respect to magnetic field strength,  
that the Zeeman effect greatly enriches the
QPO spectrum and strongly affects QPO interpretation.

\dy{Here we use the same formalism for studying the combination of axially symmetric poloidal and toroidal fields (Sec.\ \ref{s:theory}),
and apply the results to the case of purely toroidal fields (Sec.\ \ref{s:toroidal}). The latter case is simpler and can be done with more confidence}. We discuss the relation of
new theoretical results to observations
in Sec.\ \ref{s:discuss} and conclude
in Sec.\ \ref{s:conclude}.

\section{Theory}
\label{s:theory}

\subsection{General remarks}
\label{s:General remarks}

Since the standard formalism for studying magneto-elastic
oscillations in magnetars is well known (e.g., \cite{2012Gabler,2012vanHoven}), we describe it briefly. A non-oscillating star is
assumed to \dy{possess} a stationary
magnetic field $\B(\bm{r})$  
which is not too strong  
($\lesssim 5 \times 10^{15}$ G)  to violate global stellar sphericity. The 
oscillations are thought to be small, and treated in
the linear approximation as \dgy{non-dissipative} motion of incompressible matter and magnetic field perturbations
frozen into that matter. In this case, local matter elements
move along respective spherical surfaces and produce neither 
perturbations of the pressure nor generation of gravitational
waves. We focus on \dy{the} perturbations located mostly in the
magnetar crust.

According to calculations (e.g., \cite{2020KY}), the most
important place for driving the crustal oscillations is the
layer near the crust bottom,  at $\rho \sim 10^{14}$ \gcc. A 
characteristic estimate of the shear velocity there,
defined by the shear modulus $\mu$ of the crystal, is 
$v_\mu \sim \sqrt{\mu/ \rho} \sim 10^8~{\rm cm~s}^{-1}$. 
A typical estimate of the Alfv\'en velocity
gives  
$v_{ A} = {B}/{\sqrt{4 \pi \rho}} \sim 3 \times 10^7\,B_{*}
~{\rm cm~s}^{-1}$,
where $B_{*}$ is the magnetic field in units of $10^{15}$ G. 
The velocities $v_\mu$ and $v_A$ become comparable at 
$B \sim B_{\mu} \sim 3 \times 10^{15}$ G.
This is a characteristic field strength which 
separates two regimes in which the oscillations are 
mainly determined by shear effects ($B \lesssim B_\mu$)
and Alfv\'en waves ($B \gtrsim B_\mu$).
As demonstrated below, for 
high oscillation
frequencies
$B_\mu$ becomes lower. 

\subsection{Reference frames}
\label{s:frames} 

We will use three different
reference frames.

Firstly, we can employ the standard Newtonian approach 
in spherical coordinates $(r,\theta,\phi)$. It is simple and clear.

Secondly, we can describe the neutron star crust in full 
General Relativity (GR) with the
metric
\begin{equation}
ds^2 = -c^2 e^{2 \Phi} dt^2 + e^{2 \Lambda} dr^2
+ r^2 (d\theta^2 + \sin^2 \theta~ d\phi^2),
\label{e:metric}
\end{equation}
where $t$ is Schwarzschild time (for a distant observer), $r$ is circumferential radius, $\theta$ and $\phi$ are ordinary
spherical angles, while $\Phi(r)$ and $\Lambda(r)$ are two
metric functions determined by solving the standard 
GR equations for a spherical star (e.g., Ref.\ \cite{HPY2007}).

Thirdly, taking into account that the crust is thin 
and low-massive, we can also use a simplified
GR description with the metric (\ref{e:metric}),
where  $\Phi$ and $\Lambda$ 
are taken constant,
\begin{equation}
\Lambda=-\Phi, \quad e^{\Phi}=\sqrt{1-\xg}.  
\label{e:toymetric}  
\end{equation}
Here $x_g=2GM/(c^2 R)$ is the  
neutron star compactness parameter, 
$G$ is the gravitational constant, $c$ is the 
velocity of light, while $M$ and $R$
are the gravitational stellar mass and
circumferential radius, respectively. 

This approximation
is well known in neutron-star physics; it has
been used for studying various problems
of structure and evolution of the
crust (e.g., Ref.\ \cite{HPY2007}). It allows 
one to introduce a local crustal reference frame, which
is nearly flat, and then solve corresponding Newtonian problem in that frame. For comparing with full GR, one should transform the results for a distant observer. 

Note, that $dr$  in GR is a proper measure of circumferential 
distances, while proper distances in radial direction
are measured by $d \widetilde{r}=e^{\Lambda} dr$. Then,
while using the simplified GR description after starting  
from Newtonian equations in the crustal reference frame,
it would be better to treat radial derivatives of any function $f$ (designed formally as $\partial f/\partial r$) in a Newtonian
solution as 
$
 {\partial f}/{\partial \widetilde r}
\to e^{-\Lambda} \, {\partial f}/{\partial r},
$
keeping $r$ for measuring circumferential distances.
Note also that if $\widetilde{\omega}$ is an oscillation frequency in the
crustal frame, then a distant observer 
detects the
redshifted frequency $\omega = \sqrt{1 - \xg}\, \widetilde{\omega}$.

\subsection{Standard equations}
\label{s:equations}

To outline the equations which govern magneto-elastic
oscillations, we follow Refs.\ \cite{2023Yak1,2024Yak} and disregard temporarily the GR effects 
using the Newtonian approach.

The oscillations   
excite small velocities of matter elements ${\bf v}(\bm{r},t)$, small
displacements of these elements ${\bf u}(\bm{r},t)$, and small variations of magnetic fields
${\bf B}_1(\bm{r},t)$. All these quantities oscillate as $e^{{ i}{\omega} t}$, where ${\omega}$ is the angular oscillation frequency.
This overall oscillating factor in the linearized
oscillation equations can be dropped, leading to the 
stationary wave equation for small (generally complex) amplitudes $\bm{u} (\bm{r})$ and $\bm{B}_1 (\bm{r})$, and for $\omega$: 
\begin{equation}
\rho_H \omega^2 \vu= \bm{ T}_\mu + \bm{ T}_{B}.
\label{e:newton}     
\end{equation}
Here $\rho_H=\rho+P/c^2$ is the enthalpy density of
neutron star matter, a proper measure of
inertial mass density, determined by the real mass
(energy) density $\rho$ and by the pressure $P$.
Furthermore, $\bm{T}_\mu$ and $\bm{ T}_{ B}$ are the mean volumetric densities of forces
(with minis sign) determined, respectively, by the crystal elasticity
and Alfv\'en perturbations. In the case of elasticity
\begin{equation}
\dy{ T}_{\mu i}=- \frac{\partial \sigma_{ik}}{\partial x_k},\quad
\sigma_{ik}= \mu \,\left(\frac{\partial u_i}{\partial x_k} 
+\frac{\partial u_k}{\partial x_i} \right), 
\label{e:Tsigma}        
\end{equation}
$\sigma_{ik}$ being the tensor of shear deformations and $\mu$ the
shear \dy{modulus} in the isotropic-crystal approximation. In the case
of magnetic forces,
\begin{equation}
\bm{T}_{ B}= \frac{1}{4 \pi}\, \B {\bf \times }
{\rm curl}\, \B_1, \quad \B_1={\rm curl}(\vu {\bf \times} \B).
\label{e:TB}    
\end{equation}

The equations should be supplemented with boundary conditions. 
\dgy{Since local matter elements move 
along respective spheres, the crust-core interface remains
spherical. Generally, Alfv\'en perturbations can
propagate outside the crust, and one needs additional dynamical
equations in the outside magnetized regions with corresponding outside
boundary conditions. In Secs.\ D and E we argue that 
our restricted problem can be studied by solving dynamical 
equations in the crust and by requiring the radial components of shear  stresses
to vanish at crust boundaries.}

\subsection{Case of $\bm{B}=0$}
\label{s:B=0}

This case has been solved in full GR (e.g., \cite{1983ST,2007Sotani}), and we discuss these solutions.
Any stationary wave function $\vu(\bm{r})$ has two non-trivial components, $\dy{u}_\phi$
and $\dy{u}_\theta$ (since $\dy{u}_r=0$). 

It is instructive to introduce the full set of basic wave functions
for torsion oscillations
at $\B=0$ (e.g., \cite{2023Yak1}). They
can be specified  by the three quantum
numbers $(n,\ell,m)$, where $n=0,1,2,\ldots$ is the
nimber of radial nodes, $\ell$ is the orbital
quantum number which runs integer values $\ell \geq 2$
in the given problem, and 
$m$ is the azimuthal quantum number running integer values 
from $-\ell$ to $\ell$.  
The basic wave functions read
\begin{eqnarray}
u_\phi (r,\theta,\phi) & = &	r Y(r)\, 
{\rm e}^{{ i}m \phi}\, 
\frac{d\,P_{\ell}^m}{d \theta},
\label{e:uphi} \\
u_\theta (r,\theta,\phi) & = &	r Y(r)\, {\rm e}^{{ i}m \phi}\,
\frac{{i}m P_{\ell}^m}{\sin \theta},
\label{e:utheta}
\end{eqnarray}	
where $r$ is the circumferential radius,
$P_\ell^m(\cos \theta)$ is an associated Legendre
polynomial (e.g., Ref.\ \cite{1966Arfken}),
and $Y(r)=Y_{n \ell}(r)$ is a radial wave function 
obeying the equation
\begin{eqnarray}
&& Y_{,rr}+\left(\frac{4}{r}+ \Phi_{,r} - \Lambda_{,r}
+ \frac{\mu_{,r}}{\mu}\right)Y_{,r}
\nonumber \\
&  &+ \left[ \frac{{\rho}_H}{\mu} e^{-2 \Phi} \omega_\mu^2 - \frac{(\ell+2)(\ell-1)}{r^2} \right]e^{2 \Lambda}Y=0.
\label{e:Yrel}	
\end{eqnarray}
The symbols $\small{,r}$ and $\small {,rr}$ in the subscripts
mean derivatives over $r$. The frequencies of such pulsations
are denoted as $\omega_\mu$.

These oscillations are locked in the crystalline crust,
$R_1 \leq r \leq R_2$, where $R_1$ labels the boundary
between the crust and the liquid stellar core, while
$R_2$ labels the outer boundary of the crystal,
that is very close to the stellar radius $R$. At both
boundaries the radial shear stresses should vanish:
 $Y_{,r}(R_1)=Y_{,r}(R_2)=0$. Such oscillations 
are degenerate in $m$:
$\omega_\mu =\omega_{\mu n\ell}$ and $Y=Y_{n\ell}(r)$ are independent of $m$. The boundary value $Y_0=Y(R_2)$ characterizes the angular
amplitude of oscillations (in radians)  
of the outer boundary of the crystal. 
If $m=0$, crystalline matter oscillates along 
\dy{respective circles 
($u_\theta=0$, at constant $r$ and $\theta$)}; meridional motions appear at $m \neq 0$. 
Any specific stellar model affects only $Y(r)$; the
angular dependence of $\vu(\bm{r})$ stays standard.

It is important that oscillation frequencies $\omega_\mu$ can be expressed as ratios of
integrals over the crystallized matter \cite{1983ST} (also see, e.g., \cite{2020KY,2023Yak2}): 
\begin{equation}
\omega_\mu^2=\frac{\int dV\, \mu~[r^2 e^{\Phi-\Lambda} |Y_{,r}|^2
	+(\ell+2)(\ell-1)|Y|^2]}{
	\int dV \, r^2 {\rho_H}~ e^{\Lambda-\Phi} |Y|^2  },
\label{e:omegaGR}
\end{equation}
where $dV=  r^2\, d\tilde{r} \sin \theta ~ d\theta \, d\phi$ 
is a proper volume element, with $d\tilde{r}=e^\Lambda dr$. We will use this GR equation throughout
the paper for calculating $\omega_{\mu}$.

One can check, that starting from Newtonian
approach and switching to the simplified GR approach 
(\ref{e:toymetric}), one comes to the
expression which is fully
consistent with (\ref{e:omegaGR}). 
According to Ref.\ \cite{2023Yak2}, in this case one can safely
replace $\xg=2 G M/(c^2 R)$ by
$x^*_g=2GM_*/(c^2 R_*)$, where $M_*$ is the gravitational
mass enclosed in a sphere of any 
radius $R_*$ within the crust.

The functions (\ref{e:uphi}) and (\ref{e:utheta}) will be used as basic wave functions for studying magneto-elastic oscillations. 

\subsection{First-order perturbation theory}
\label{s:FOPT}

Following Refs.\ \cite{2009SE,2023Yak1,2024Yak} 
we continue studying magneto-elastic oscillations
using the first-order 
perturbation theory (FOPT) where the magnetic term
$\bm{T}_B$ in Eq.\ (\ref{e:newton}) is treated as a
perturbation. We begin with the plain Newtonian approximation.

For simplicity, we study the $\B$ field configurations which are 
axially symmetric
with respect to the magnetic axis $z$,
\begin{equation}
B_r=B_r(r,\theta),\quad B_\theta=B_\theta(r,\theta),
\quad B_\phi=B_\phi(r,\theta).
\label{e:axialsymmetry}   
\end{equation}
The field
components $B_r$, $B_\theta$ and $B_\phi$ 
can be arbitrary functions of $r$ and $\theta$ satisfying magnetic flux conservation; $B_r$ and $B_\theta$ are
the poloidal field components, while $B_\phi$ is the toroidal
one. We will also impose the mirror symmetry 
of $\B (\bm{r})$-lines with
respect to the magnetic equator.

Our zero-order solutions are those described in Sec.\ \ref{s:B=0}. Zero-order wave functions
 $\vu$ are localized in the
crust, and oscillation frequencies  $\omega_{\mu}$
are degenerate in $m$.
A first-order correction to $\omega_\mu$
will break the degeneracy
owing to the Zeeman effect. 
The problem is
similar to finding a first-order correction to a non-perturbed degenerate
energy level of a quantum system (e.g., Ref.\ \cite{LL76}). 
In case of not too strong Zeeman splitting, it is sufficient to use 
($2 \ell+1$) zero-order wave functions $\vu$ corresponding to
a zero-order $\omega_{\mu}$ at given $n$ and $\ell$, and calculate then the perturbation matrix ${T}_{m'm}$
[of dimension $(2 \ell+1) \times
(2 \ell+1)$] 
on this restricted basis of zero-order states:
\begin{equation}
{T}_{m' m}= \int_{\rm crust} d V \, \vu_{n \ell m'}^*\, \bm{T}_B(\vu_{n \ell m}).
\label{e:Tmatrix}    
\end{equation}
This matrix naturally occurs \cite{2023Yak1}, if we \dy{multiply} Eq.\ (\ref{e:newton}) by $\vu^*$ and integrate over the
star.
Since zero-order states are confined in the crust, the integration is restricted by the
crust alone. One can easily check that, for the chosen magnetic
field configuration (\ref{e:axialsymmetry}), the 
matrix $T_{m'm}$ is diagonal due to integration over $\phi$ from 0 to
$2 \pi$. This means that the magnetic
interaction does not mix the states with different $m$,
so that
labels $(n, \ell, m)$ retain their previous meaning 
(as in Sec. \ref{s:B=0}). Then 
any first-order `magnetically corrected' oscillation frequency 
with fixed $n$ and $\ell$ is given by the sum rule \cite{2023Yak1}:
\begin{equation}
\omega^2=\omega_{\mu}^2+\omega_{B}^2,
\label{e:omegaB1}	
\end{equation}	
where
\begin{equation}
\omega_\mu^2= 
\frac{\int d V\, \vu^* \bm{ T}_\mu}{\int d V \, \rho_H |\bm{u}^2|}, \quad
\omega_B^2= 
\frac{\int d V\, \vu^* \bm{T}_B}{\int d V \, \rho_H |\bm{u}^2|}.
\label{e:omega-mu-B}	   
\end{equation}   

Hence, the true oscillation frequency $\omega$ is formally expressed via two
frequencies, $\omega_\mu$ and
$\omega_B$, and $\omega_{\mu}$
can be expressed as (\ref{e:omegaGR}). 

Under the formulated assumptions the sum rule (\ref{e:omegaB1})
is exact. It has to be true in full GR.

Since $\omega_\mu$ is independent of $B$,
all magnetic effects, including
\dy{the} Zeeman splitting, are incorporated in $\omega_B$. The
derivation of (\ref{e:omegaB1}) suggests that $\omega_B$ is
smaller than $\omega_\mu$, but we will
extend Eq.\ (\ref{e:omegaB1}) 
to higher $B$.  We will present additional 
arguments in favor for this extension in Sec.\ \ref{s:toroidal}.

The ratios of integrals in Eqs.\
(\ref{e:omega-mu-B}) are of similar
origin. 
The quantity $\omega_B^2$ 
can be rewritten as
\begin{equation}
\omega_B^2= \dgy{
\frac{\int d V\,  J_B}{\int d V \, \rho_H |\bm{u}^2|}},
\quad
\dgy{J_B=\frac{1}{2\pi} \int_0^{2 \pi} d\phi \,{\vu^* \bm{T}_B}}.
\label{e:def-IB}
\end{equation} 
\dgy{Here $J_B=J_B(r,\theta)$ is an effective `energy density' $\vu^* \bm{T}_B$ 
in a matter element at fixed $r$ and $\theta$ averaged
over $\phi$ (or, equivalently, over pulsation period if we formulated
time-dependent perturbation theory). 
The averaging can be done analytically and separately
because the dependence of $\vu^* \bm{T}_B$ 
on $\phi$ in the assumed
axial symmetry of $\B$ is standard. This greatly simplifies
the result because many $\vu^* \bm{T}_B$ terms are canceled out.}
Taking $\bm{T}_B$ from
Eq.\ (\ref{e:TB}), the magnetic field configuration from Eq.\ (\ref{e:axialsymmetry}), and using the magnetic flux conservations, by direct calculation we obtain another sum rule
\begin{equation}
    I_B \equiv 4 \pi J_B =I_{Bp}+I_{Bt}, 
\label{e:BpBt}    
\end{equation}
where the terms  $I_{Bp}$ and $I_{Bt}$ contain,
respectively, only poloidal and toroidal field components.

It is convenient to
write $I_{Bp}=I_{Bp0}+I_{Bp1}$. 
The term $I_{Bp0}$ was obtained in
\cite{2023Yak1} for describing fundamental oscillations ($n=0$, $Y= Y_0$, see Sec.\ 
\ref{s:fundamental modes}): 
\begin{eqnarray}
I_{Bp1}& = & (B_\theta \, B_{\theta,\theta} \cot \theta
+B_r B_\theta \cot \theta +r B_r B_{\theta ,r}\ \cot \theta
\nonumber \\
& -& B_\theta^2) \,
P_{,\theta}^2 \,|Y|^2
-(B_\theta \, B_{\theta,\theta}+B_r B_\theta+ r B_r B_{\theta ,r})
P_{,\theta}	P_{,\theta \theta} 
\nonumber \\
&\times &|Y|^2  - B_\theta^2 P_{,\theta} P_{,\theta \theta \theta }\,|Y|^2 
  + [  B_\theta^2 P_{,\theta}^2 - (B_\theta B_{\theta,\theta} 
\nonumber \\  
 & + & B_\theta^2 \cot \theta+ 2 B_r B_{r,\theta}
 + B_r B_\theta + r B_r B_{\theta, r}) P P_{,\theta}    
\nonumber \\
&+&
B_r (B_{\theta,\theta}+ B_\theta \cot \theta
+r B_{\theta,r\theta}+ r B_{\theta, r} \cot \theta
\nonumber \\
&- &B_{r,\theta \theta} + B_{r,\theta} \cot \theta)\,P^2]~ |Y|^2 \,(m/\sin \theta)^2.
\label{e:IBp0}
\end{eqnarray}
Indices $\theta$ and $r$ after commas in the subscripts again denote
respective partial derivatives; $P=P_{\ell}^{m}(\cos \theta)$. 
 
The second term $I_{Bp1}$ reads
\begin{eqnarray}
 I_{Bp1}& = & r Y^* P_{,\theta}B_r [Y_{,r}P_{,\theta} (B_\theta \cot \theta -2 B_r -r B_{r,r} )
 \nonumber \\
 &- & Y_{,r} B_\theta P_{,\theta \theta} - rY_{,rr} B_r P_{,\theta}]
 -r Y^* Y_{,r}B_\theta P_{,\theta}
 \nonumber \\
 & \times &
 [P_{,\theta}B_r \cot \theta +P_{,\theta} B_{r,\theta}
 +P_{,\theta \theta}B_r
 \nonumber \\
  &-& (m/\sin \theta)^2 P B_r]
  -r Y^* P B_r \,(m/\sin \theta)^2
  \nonumber \\
  &\times & [P_{,\theta}B_{\theta}  Y_{,r}-P (B_{\theta,\theta} + B_\theta \cot \theta -2 B_r) Y_{,r}
 \nonumber \\
 &+ & r P B_r Y_{,rr}+r P B_{r,r} Y_{,r}].
\label{e:IBp1}	
\end{eqnarray}
It is needed for studying oscillations with radial nodes
($n>0$).

Finally, in the presence of toroidal field
we obtain
\begin{eqnarray}
 I_{Bt}&=& |Y|^2 P B_\phi \, (m/ \sin \theta)^2 [ (m /\sin \theta)^2 P B_\phi 
\nonumber \\
& - & P_{,\theta \theta} B_\phi - 2 P_{,\theta} B_{\phi,\theta}
+ P_{,\theta} B_\phi \cot \theta
\nonumber \\
&+& P B_{\phi,\theta} \cot \theta -P B_{\phi,\theta\theta} -P B_\phi/ (\sin \theta)^2 ].
\label{e:IBt} 
\end{eqnarray}  

Equations (\ref{e:IBp0})--(\ref{e:IBt}) make the integration
over $dV$ in the nominator of (\ref{e:def-IB}) quite feasible.
It is reduced to \dy{the} integrations over $\theta$ (from 0 to $\pi$) and $r$ (from $R_1$ to $R_2$). The integration in the denominator is also simplified, as detailed in 
\cite{2023Yak1}. 

Now we return from the Newtonian approach 
to the simplified GR approach
 (Sec.\ \ref{s:frames}). 
As a result, Eq.\ (\ref{e:def-IB}) takes
the same form as in \cite{2023Yak1}:
\begin{eqnarray}
\omega_{B}^2& =& \frac{(1-\xg) \int_{R_1}^{R_2} d r \,r^2
	\int_0^{\pi} \sin \theta \, d \theta \, I_B}
{4 \pi \, \dgy{
	\int_{R_1}^{R_2} d r \, r^2 \rho_H \int_0^{\pi} \sin \theta \, d \theta \,|\bm{u}|^2  } } 
\nonumber
 \\
&=&
\frac{(1-x_g)  \int_{\rm crust} d V \, I_B}
{4 \pi \Xi(\ell,m)\,
	\int_{\rm crust} d V \, |Y|^2 \rho_H r^2 }, 
\label{e:omegaB4}	
\end{eqnarray}
where $dV$ is the same as in Eq.\ (\ref{e:omegaGR})
(\dy{including angular integration}, with constant $\Lambda$), and
the normalization coefficient $\Xi(\ell,m)$ reads \dy{\cite{2023Yak1}}
\begin{equation}
\Xi(\ell,m)= \frac{2 \ell (\ell+1) (\ell+m)!}{(2 \ell+1)(\ell-m)!}.
\label{e:Xi}
\end{equation} 
The factor $(1-\xg)$ accounts for gravitational
redshift of the squared 
oscillation frequency, and $r$ is 
circumferential radius. The field 
$\B(\bm{r})$ is treated 
as the magnetic field in the crustal reference frame.
While calculating $I_B$ from Eqs.\ (\ref{e:IBp0})--(\ref{e:IBt}), it is desirable to
calculate radial derivatives as prescribed in
Sec.\ \ref{s:frames}. 
This was not done in our previous work (Refs. \cite{2023Yak1,2024Yak}). This is not an error, but rather
a trick to be more consistent with GR.  
   
Note that the function $Y(r)$ was not included in $I_B$ and in the
denominator of the expression for
$\omega_B^2$ in Ref.\ \cite{2023Yak1}. That paper was devoted 
to fundamental oscillations in which case $Y(r)$ was set constant (see Sec.\ \ref{s:fundamental modes}) and  naturally dropped in the ratio of integrals.

So far we have not used the assumed symmetry 
of $\B$-field lines
with respect to the magnetic equator. Imposing this
assumption, we evidently obtain that 
oscillation frequencies with $m \neq 0$ and opposite $m$ become equal. 
Accordingly, the frequencies can be labeled by 
the index $m$ which runs \dy{the} integer values $m=0,1,\dots \ell$.
The frequencies with $m=0$ are non-degenerate,
\dy{while those with $m>0$ are now degenerate twice}.

\subsection{Alfv\'en wave problem}
\label{s:alfven}

Calculations in Sec.\ \ref{s:FOPT} have been
performed using the FOPT 
with respect to the magnetic field strength. It is expected 
to be accurate for describing \dy{the} Zeeman splitting at not too
high magnetic fields. Its firm applicability
conditions deserves further attention, particularly, in
view of the Alfv\'en wave problem. 

The theory assumes vanishing elastic shear stresses
at the boundaries of crystallized matter. Nevertheless, in the presence of Alfv\'en perturbations, there
are additional magnetic stresses, which should 
appear in higher-order versions of the perturbation
theory. Alfv\'en perturbations
can spread magneto-elastic 
oscillations outside the crust. This important 
effect has been studied in many publications, 
e.g., \cite{2006Levin,2006Glampeda,2007Levin,2009Colaiuda,2009CD,2011Gabler,2011vanHoven,2011Colaiuda,2012vanHoven}.
These studies have been mainly restricted to 
axially symmetric perturbations which drive 
the oscillations only with $m=0$ (see Sec.\ \ref{s:toroidal}). More work is required to explore all
consequences of this effect.

\section{Toroidal magnetic fields}
\label{s:toroidal}

\subsection{Squared magnetic frequency $\omega_B^2$}
\label{s:toroidal start}

Here we focus on the case of pure 
toroidal magnetic field $\B$ determined
by $B_\phi(r,\theta)$ in Eq.\ (\ref{e:axialsymmetry}). This case has
attracted little attention, but 
it is important and very
simple. To avoid discussing the Alfv\'en wave
problem (Sec.\ \ref{s:alfven}), we restrict ourselves to the field configurations confined fully in
the solid crust. The magneto-elastic
oscillations cannot then spread outside. Such toroidal
fields should be supported by poloidal electric
currents located in the crust.
In this case in Eq.\ (\ref{e:omegaB4}) we have
$I_B=I_{Bt}$.

By way of illustration, we 
consider a toroidal magnetic \dy{field} of the form 
\begin{equation}
	B_\phi= B_0\, {\sin \theta} \,
	\psi(x),\quad x=\frac{R_2-r}{R_2-R_1}.
\label{e:B_phi}	
\end{equation}
Here $B_0$ is  the maximum field
in the crust. The
dependence of $B_\phi$ on $r$ is
now determined by a dimensionless function
$\psi(x)$ of  
a dimensionless depth $x$ that
varies from $x=0$ at the outer surface of 
the crystalline matter ($r=R_2$) to $x=1$ at the
crust bottom ($r=R_1$). The 
function $r\,\psi(x)$ 
is often called the source function (e.g. \cite{2008Aguilera}). We 
assume that
$\psi \to 0$ as $x \to 0$ and
$x \to 1$ to avoid 
electric currents at both
boundary surfaces. The maximum 
value $\psi=1$ occurs 
at the depth, where $B_\phi$ reaches its maximum. Note that
in our calculations $R_2$ has been taken
only slightly
lower than $R$, so that the difference
can be safely neglected in final expressions.

The dependence of $B_\phi$ on $\theta$ in Eq.\
(\ref{e:B_phi}) 
is fixed. Then $B_{\phi,\theta}=
B_\phi \cot \theta$ and $B_{\phi, \theta
\theta}=-B_\phi$, and Eq.\ (\ref{e:IBt})
reduces to
\begin{equation}
	I_{Bt} = |Y|^2 \ell (\ell+1)
   \frac{m^2 P^2  B_\phi^2}{\sin^2 \theta},
\label{e:IBt2}   	 
\end{equation}	
where the differential equation 
for $P=P_{\ell}^m (\cos \theta)$ has
been used (e.g., \cite{1966Arfken}). Now the integration over
$\theta$ in the nominator of (\ref{e:omegaB4}) is carried out
analytically, and we come to
the very
simple expression 
\begin{equation}
		\omega_{B}^2=
	(1- x_{ g})\, \frac{ m^2 \dy{B_0^2} \int_{R_1}^{R_2} 
		dr \, \psi^2 |Y|^2 r^2}
	{4 \pi 
		\int_{R_1}^{R_2}  dr \, 
		{\rho_H}\,|Y|^2 r^4 } .
	\label{e:omegaB6}	
\end{equation}
It shows \dy{the} Zeeman splitting of 
oscillation frequencies $\omega$, in which
$\omega_B^2 \propto m^2$ and the Legendre 
polynomials $P_\ell^m (\cos \theta)$
are integrated out. At $m=0$ we have
$\omega_B=0$, and the total
oscillation frequency in Eq.\ (\ref{e:omegaB1}) becomes
$\omega=\omega_\mu$, being   
independent of $B$. 
In
Eq.\ (\ref{e:omegaB6}) we are left with
the ratio of two simple radial integrals which can be easily computed for any given model of the neutron star crust and any meaningful
function $\psi(r)$.  

\subsection{Neutron star models}
\label{s:models}

For illustration, we use neutron star models
based on one equation of state (EOS) of dense matter
called the BSk21 EOS. It belongs to 
a family of well elaborated Brussel-Montreal EOSs. They are unified: energy-density 
functionals, employed for constructing the EOSs in the crust and the core, are the
same. The core is assumed to contain
nucleons, electrons and muons. The BSk21 crust consists
of electrons and spherical atomic nuclei. In deep crustal
layers ($\rho \geq 4.28 \times 10^{11}$ \gcc) there appear free neutrons, and near the crust-core interface ($\rho_{\rm cc} \approx 
1.34 \times 10^{14}$ \gcc) there occurs an admixture of free protons. 

Various properties of neutron stars with the BSk21 EOS are described by convenient analytic approximations in Ref.\ \cite{BSk2013}. The maximum mass of the neutron star mass is
2.27~$\msun$; its radius $R=11.25$ km. The neutron star
with the canonical mass of $1.4\, \msun$ has radius 
$R=12.60$~km, and the core radius $R_1=11.55$~km. A lighter
$1$ $\msun$ star has $R=12.48$ km and
$R_1=10.92$ km. A very massive 
$2.2$ $\msun$ star has $R=11.81$ km and $R_1=11.39$ km.
We will begin with the $1.4 \msun$ model and then extend the results
over the $(1-2.2)\, \msun$ mass range. 

We will use the same microphysics of the crust
as in Refs.\ \cite{2023Yak2,2023Yak1,2024Yak}.

\subsection{Fundamental modes ($n=0$): Theory}
\label{s:fundamental modes}

Here, using FOPT, we study fundamental oscillation modes,
with nodeless ($n=0$) radial wave function $Y(r)$. 
This case
is important because it corresponds to 
magnetar QPOs with  frequencies
lower than a few hundred Hz.  Corresponding 
wave functions are known (e.g. Ref.\ \cite{2020KY})
to be nearly independent of
$r$, and can be approximated by a constant,
$Y \to Y_0$. This
constant drops out in the ratios
of integrals in all the expressions for
$\omega_\mu$ and $\omega_B$, e.g., in Eqs.\ (\ref{e:omegaGR}) and (\ref{e:omegaB6}).

Then the exact GR equation (\ref{e:omegaGR})
becomes
\begin{equation}
\omega_\mu^2={\textstyle{\frac{1}{4}}
	(\ell+2)(\ell-1)} \, \omega_{\mu 0}^2 ,
\label{e:omegaGRfund}
\end{equation}
where $\omega_{\mu0}$ is the frequency of the lowest 
torsion fundamental mode ($\ell=2$),
\begin{equation}
\omega_{\mu 0}^2=   \frac{4 \int_{\rm crust} dV\, \mu}{
	\int_{\rm crust} dV \,r^2 {\rho_H}~ e^{\Lambda-\Phi} }	
=(1-x_g) \frac{8 W_\mu}{3 I_0 }.
\label{e:omegamu0}
\end{equation}
The last expression is
obtained by employing 
our simplified GR model of the crust (Sec.\ \ref{s:frames}), in which $e^{\Phi-\Lambda} \to (1-x_g)$.
Accordingly,
$\omega_{\mu 0}^2$ is determined by
$I_0$, which is the \dy{standard} moment of inertia of the crust in the crustal
reference frame, and by $W_\mu$, which  is the shear modulus $\mu$ integrated 
over the crust in the same frame. In calculations, we use the well known
expression $\mu=0.1194\,n_i\,Z^2 e^2/a_i$ \cite{1990Ogata},
where $n_i$ is the number density of atomic nuclei, $Ze$ 
is the electric charge of one nucleus, 
and $a_i=(4 \pi n_i/3)^{-1/3}$ 
is the ion-sphere radius. Since the Coulomb binding energy per
one nucleus is $\dy{\approx} 0.9 Z^2e^2 /a_i$, we have
$W_\mu=0.133 \, W_C$, $W_C$ being the Coulomb energy of the
crust. 

As for the squared magnetic frequency 
$\omega_B$, given by 
(\ref{e:omegaB6}), it can be written as
\begin{equation}
\omega_{B}^2=m^2 \omega_{B0}^2, \quad \omega_{B0}^2=
(1- x_{ g})\, \frac{2 W_B}{ I_0} 
\label{e:omegaB7}	
\end{equation}
where 
\begin{equation} 
	W_B= \int_{\rm crust} dV 
	\frac{B_\phi^2}{8 \pi} \,
	=
	\frac{\kappa}{3} B_0^2 R_2^2
	(R_2-R_1),
	\label{e:W_B}
\end{equation}
is the magnetic energy of the crust \dy{and}
\begin{equation}
 \kappa=\int_0^1 (r/R_2)^2 \psi^2(x)\, dx,
 \label{e:kappa}
\end{equation}
\dy{is} a dimensionless numerical coefficient.
This coefficient depends mostly
on $\psi(x)$. Also, it slightly depends    
on a stellar model because
of the factor $(r/R_2)^2$ which slowly decreases
from its maximum value 1 at $x=0$ to $(R_1/R_2)^2$ at
the crust bottom. \dy{For $B_\phi$ configurations different
from (\ref{e:B_phi}) a numerical
factor in (\ref{e:omegaB7}) will differ from 2.}

According to Eqs.\ (\ref{e:omegamu0}) and (\ref{e:omegaB7}),
physical meanings of $\omega_{\mu0}$ and $\omega_{B0}$, are
similar: while the former is determined by the Coulomb
energy of the crust, the latter is determined by the
magnetic energy.

To be specific, we will mainly use
\begin{equation}
	\psi_0(x)=16 x^2 (1-x)^2, 
\label{e:psi0}	
\end{equation}
although we will try other $\psi(x)$
which renormalize $\kappa$. Neglecting
the factor  $(r/R_2)^2\approx 1$ in (\ref{e:kappa}),
we obtain an estimate $\kappa_0 \approx 128/315=0.40635$.

Now we can rewrite Eq.\ (\ref{e:omegaB7}) as
\begin{equation}
\omega_{B}^2=m^2 \omega_{B0_*}^2 B_*^2, \quad B_*=
\frac{B_0}{10^{15}\,{\rm G}},
\label{e:omegaB7a}	
\end{equation}
where $\omega_{B0_*}$ is a convenient measure 
of $\omega_B$ at a typical value of $B_0=10^{15}$ G; $\omega_{B0_*}$ 
is independent of $\ell$, $m$
as well of a specific value of $B_0$, 
but depends on $\psi(x)$.  

Then the frequency of any fundamental magneto-elastic oscillation
mode ($n=0$, any allowable $\ell$ and $m$), as
calculated in FOPT, 
is determined by the two
auxiliary frequencies, $\omega_{\mu0}$ and 
$\omega_{B0*}$, one and the same for a given stellar model. Note that $\omega_{B0} \propto B_0 \sqrt{\kappa}$. 

In cyclic 
frequencies $\nu=\omega/(2 \pi)$ we
have
\begin{equation}
 \nu_{0 \ell m}^2={\textstyle \frac{1}{4}}(\ell+2)(\ell-1)\nu_{\mu 0}^2+
        m^2 \nu_{B0_*}^2 B_*^2,
\label{e:nu0} 
\end{equation}
with the two constants, $\nu_{\mu0}$ and $\nu_{B0_*}$.  

As discussed in Refs.\ \cite{2020KY,2023Yak2}, the simplistic 
$\ell$-dependence  of (\ref{e:omegamu0}) and
(\ref{e:nu0}) can be violated at $\ell \gg 10$. Equation 
(\ref{e:nu0}) can also become
inaccurate at large $m$ but this is not vitally important
for applications (see Sec.\ \ref{s:discuss}).

\subsection{Fundamental modes for a $1.4\, \msun$ star}
\label{s:n014msun} 

Here we discuss fundamental magneto-elastic oscillation frequencies calculated for a $1.4\, \msun$ neutron star model
with the BSK21 EOS. 

We begin with the FOPT results assuming $\psi(x)=\psi_0(x)$, see Eq.\ (\ref{e:psi0}). The frequencies
are given by Eq.\ (\ref{e:nu0}), being determined by
the two numbers, $\nu_{\mu 0}$ and $\nu_{B0*}$, listed in Table \ref{tab:main1}.

\renewcommand{\arraystretch}{1.2}
\begin{table}[ht]
	\caption{\label{tab:main1}
		Eight auxiliary constant frequencies (in Hz), which determine magneto-elastic oscillation
		spectrum (different $\ell$ and $m$)  
		of fundamental modes ($n=0$, first two values), and ordinary modes (Sec.\ \ref{s:freqn1}) with $n=1$ (next three
		values) and $n=2$ (last three values)
		for a 1.4 $\msun$ neutron star with the BSk21
		EOS.}
	\begin{ruledtabular}		
		\begin{tabular}{ c c c c c c c c   } 
			$\nu_{\mu 20}$ & $\nu_{B0*}$ & $\nu_{\mu 1}$ &  $\delta \nu_{\mu 1} $  &  $\delta \nu_{B1*}$ & $\nu_{\mu 2}$ &  $\delta \nu_{\mu 2}  $  & $\nu_{B2*} $ \\  
			\colrule 
			23.06 & 
			4.192 & 829.7  & 12.98  & 15.98 & 1327  & 14.87  & 22.96
			\\   	       	    		
		\end{tabular}
	\end{ruledtabular}	
\end{table}
\renewcommand{\arraystretch}{1.0} 

\begin{figure}[t]
	\includegraphics[width=0.5\textwidth]{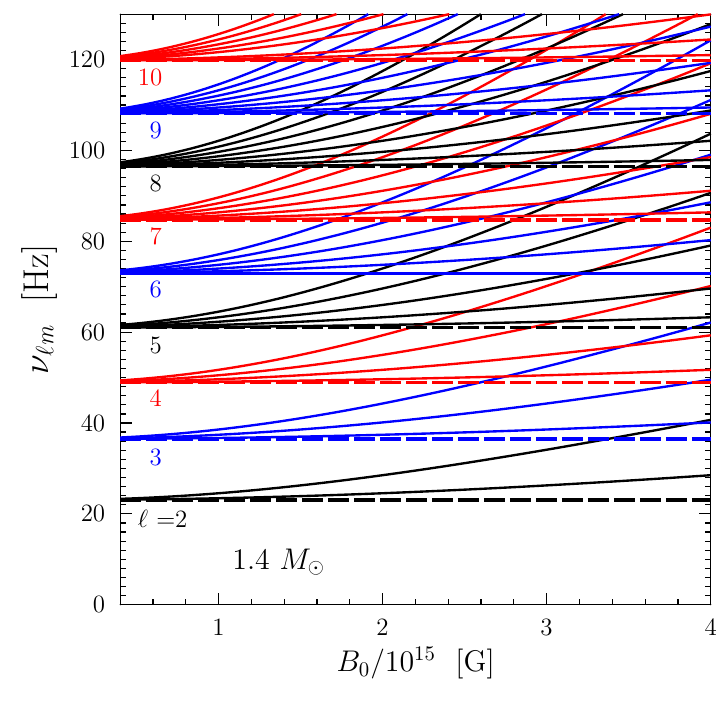}%
	\caption{\label{f:freqn0} 
		Frequencies of fundamental ($n=0$) magneto-elastic
		oscillations, as calculated
		using FOPT, versus
		maximum toroidal 
		magnetic field $B_0$
		in the crust for a 	$1.4 \msun$
		neutron star
		model (BSk21 EOS). Shown are
		the bunches of modes with
		$\ell$ from 2 to 10, and
		$m$ from 0 to $\ell$
		in each branch. The frequencies
		with $m=0$ (dashed lines) are independent
		of $B$. 
	}
\end{figure}

Figure \ref{f:freqn0} presents
the frequencies $\nu_{0\ell m}$ for $\ell=2,\dots,10$ and
all allowable $m$ at $\nu \leq 130$ Hz as a
function of $B_0$. One can see nine bunches of $\nu(B_0)$ curves.
Each bunch corresponds to a fixed $\ell$ and contains 
$\ell+1$ curves (with $m$ from 0 to $\ell$) plotted in one color.
The lowest (dashed) curve in a bunch refers to $m=0$. In our
approximation, it is independent of $B_0$, being solely
determined by the elastic shear, $\nu=\nu_\mu$. With increasing
$m$, other (solid) curves in a bunch grow up, demonstrating
Zeeman splitting. Neglecting the Zeeman effect leads to the loss of all oscillation modes with $m>0$.

As long as $B_0 \ll 10^{14}$ G, the spitting is almost
invisible, but
it becomes more pronounced  with increasing $B_0$.
With increasing $\ell$, the number of Zeeman components in a bunch grows up and the Zeeman splitting becomes  
visible at lower $B_0$. 

It is important to mention crossings of pairs of modes from
different bunches at sufficiently high $B_0$. Such
crossings turn usually to well known quasi-crossings due
to resonance interactions of the modes (see, e.g.\ Ref.\ \cite{LL76}). These effects 
are not described by our FOPT approach: one needs to 
include higher-order perturbation terms. This 
can be an interesting problem for the future.

\begin{figure}[t]
	\includegraphics[width=0.5\textwidth]{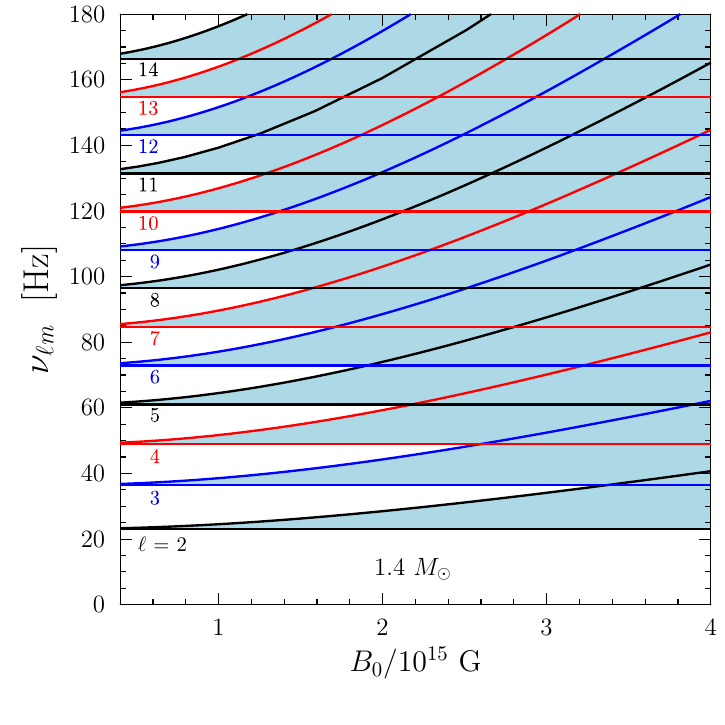}%
	\caption{\label{f:freqn0a} 
		Same as in Fig.\ \ref{f:freqn0}
		in a sketchy manner but for higher
		number of bunches ($\ell \leq 14$). Frequencies in each bunch fall in  
		darkened space between lowest
		($m=0$) and highest ($m=\ell$)
		frequencies. See the text for details.
	}
\end{figure}

Figure \ref{f:freqn0a} is a sketch of the same picture, as
in Fig.\ \ref{f:freqn0}, but extended to higher frequencies by including additional bunches 
with $\ell$ from 11 to 14.
To avoid plotting over-dense Zeeman
splittings at high $\ell$, each bunch is specified by 
the frequencies of the lowest ($m=0$) and highest 
($m=\ell$) modes drawn in one color. A space between these limiting
frequencies is darkened. Roughly
speaking, in the $\nu - B_0$ plane 
one can distinguish 
domains of two types: forbidden (white) and 
allowed (darkened) ones. The region of rather
low $\nu$ and $B_0$, where neighboring bunches do not cross,
contains a lot of white domains which
are forbidden for oscillations 
of a given star. The darkened domains at higher $\nu$ and $B_0$ are densely filled with Zeeman structures. If observed oscillation frequencies fall into these domains, it would 
easy to explain them for the given star.
One needs $B_0 \gtrsim 10^{15}$ G for reaching these `allowable'
domains at $\nu \gtrsim 140$ Hz, but  $B_0 \gtrsim 3 \times 10^{15}$ G  at $\nu \lesssim 140$ Hz.

\begin{figure}[th]
	\includegraphics[width=0.45\textwidth]{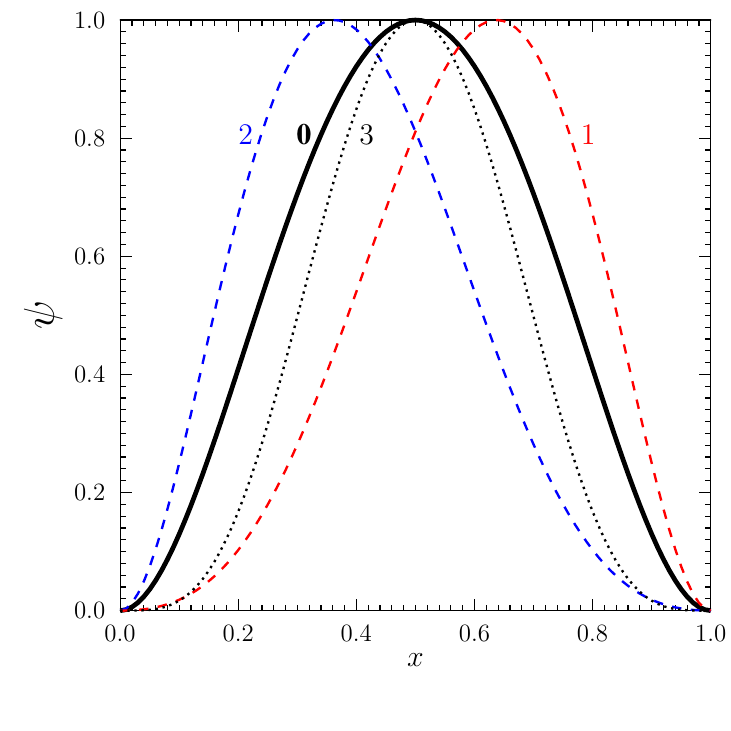}
	\caption{\label{f:bfun} Four versions of the function $\psi(x)$, that determines the 
	radial behavior of $B_\phi$, versus dimensionless 
	depth $x$ within the crystalline layer. The thick line 0 is the basic version, Eq.\ (\ref{e:psi0}). Other versions 1--3 
	are for
	checking sensitivity of oscillation frequencies to
	the shape of $B_\phi(r)$ profile.}
\end{figure}

\begin{figure}[th]
	\includegraphics[width=0.45\textwidth]{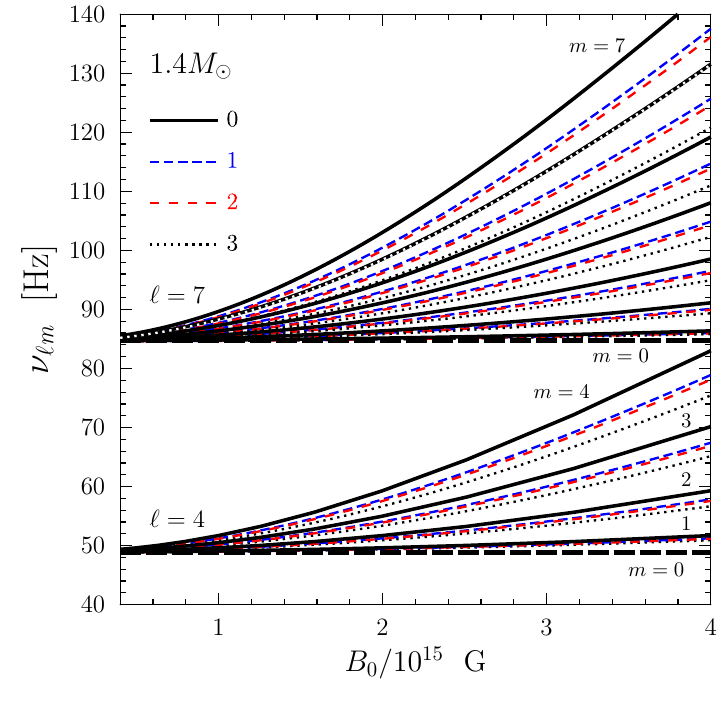} 
	\caption{\label{f:freqbfun} 
		Sensitivity of magneto-elastic
		oscillation frequencies to 
		the radial dependence of $B_\phi$. We show two bunches
		of oscillation modes, with  
		$\ell$=4 and 7, versus $B_0$
		for four versions of the function $\psi$ in Fig.\
		\ref{f:bfun}. See the text for details.
	}
\end{figure}

So far we have studied only one radial profile (\ref{e:psi0}) of $B_\phi(r)$ through the neutron star crust. In Fig.\ \ref{f:bfun} we plot this profile by the thick black
line (line 0). This profile profile
is centered at $x=1/2$, just in the middle between the
inner and outer boundaries, $R_1$ and $R_2$. For
exploring the sensitivity of our results to the profile shape,
we have considered three more $\psi(x)$ models
plotted in Fig.\ \ref{f:bfun} by thinner lines 1, 2, and 3.
 The profiles
1 and 2 are slightly narrower; the peak of   profile 1
is shifted to the surface, while the peak of 
profile 2 
is shifted to the stellar core. The shifts are 
symmetric with respect to $x=1/2$. The profile 4 is again
centered at $x=1/2$ being noticeably narrower than other profiles. Actually,
we just set $\psi_4(x)=\psi_0^2(x)$.

Corresponding oscillation
frequencies are compared in Fig.\ \ref{f:freqbfun}. For clarity,
we plot only two oscillation bunches,
for $\ell=4$ and $7$.

The thick black lines refer to our basic
model; they are the same as in Fig.\ \ref{f:freqn0}. Other lines in Fig.\ \ref{f:freqbfun} refer to models
1, 2, and 3. Changing  $\psi(x)$ 
renormalizes the frequency $\nu_{B0*}$ in
Eq.\ (\ref{e:nu0}). Instead of $\nu_{B0*}=$4.19 Hz for
model 0 in Table \ref{tab:main1}, we would have 3.87 Hz, 3.81 Hz and
3.59 Hz, for models 1, 2 and 3, respectively.
The renormalized   $\nu_{B0*}$ for models 1 and 2, as well as  respective oscillation spectra in
Fig.\ \ref{f:freqbfun}, 
appear to be very close. The
frequencies for model 1 are slightly
higher because of somewhat higher  
crustal magnetic energy $W_B$ [see Eq.\ (\ref{e:W_B})]
due to the factor $(r/R_2)^2$
under the integral in (\ref{e:kappa}). The lower $W_B$, the weaker magnetic field effect on oscillation frequencies.

Note a funny effect of frequency degeneracy in
Fig.\ \ref{f:freqbfun}: accidentally, the
oscillation frequency at $\ell=7$ and $m=6$ 
for the basic model 0 coincides with
the frequency at $\ell=7$ and $m=7$ for
model 4, and the two curves merge. Similar degeneracies
are discussed below in Sec.\ \ref{s:discuss}. 

\subsection{Restricted variational estimates}
\label{s:variational}

\begin{figure}[th]
	\includegraphics[width=0.5\textwidth]{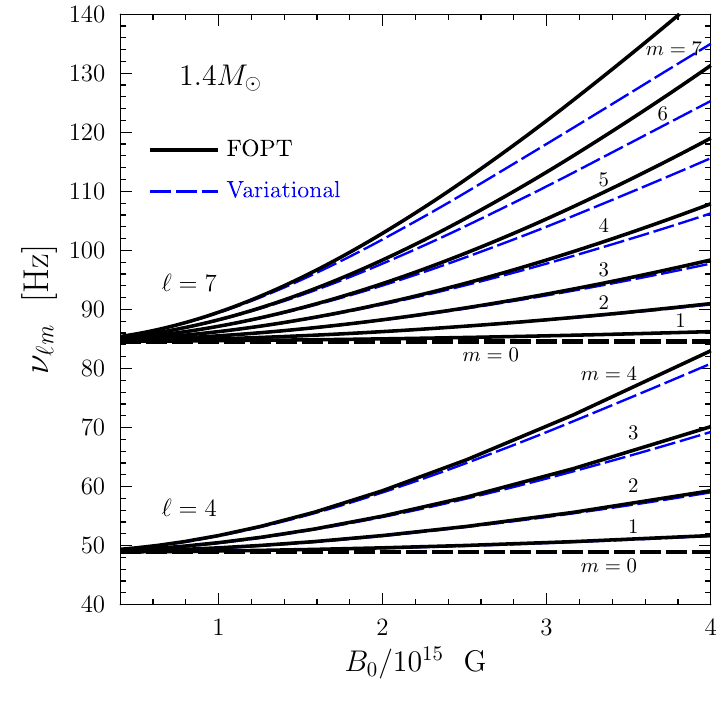} 
	\caption{\label{f:freqvar} 
		Comparing oscillations 
		frequencies calculated using
		FOPT with variational estimates. Shown are magneto-elastic
		oscillation frequencies 
		versus $B_0$ for two
		bunches of modes with
		$\ell$=4 and 7 at various
		$m$.}
\end{figure}

So far we have presented only FOPT calculations
of magneto-elastic oscillation frequencies.
Here we check these results using restricted
variational estimates. To this aim, we 
use the expression $\nu^2 \{ Y \} = \nu^2_\kappa \{ Y \}+\nu^2_B \{ Y \}$ as
a functional of the radial wave function $Y(r)$. In the FOPT, this wave function is
constant, $Y=Y_0$, and drops out of the theory. Using $Y(r)$ as a variational function for minimizing  $\nu^2 \{ Y \}$
allows us to 
estimate the validity of FOPT and possible
deviations of $\nu$ from the FOPT values.

For  
studying the fundamental oscillations, the
variational function $Y(r)$ should have no nodes at
$R_1 \leq r \leq R_2$. Also, it should satisfy
proper boundary conditions ($Y_{,r}(r)=0$
at $r=R_1$ and $R_2$). The functionals 
$\nu^2_\kappa \{Y\}$ and $\nu^2_B \{Y \}$ are
given by Eqs.\ (\ref{e:omegaGR}) and (\ref{e:omegaB6}), respectively. As before, we employ
our $1.4 \, \msun$ stellar model. The radial dependence
of $B_\phi$ is given by
Eq.\ (\ref{e:psi0}). The variational function
has been chosen in the form
\begin{equation}
	Y(r)= \left[  1 + \dy{w} \sin^2 ( \pi x / 2) \right] Y_0,
\label{e:variational}	
\end{equation}
where $x$ is the same dimensionless depth
as in Eq.\ (\ref{e:B_phi}), and $\dy{w}$ is a single
variational parameter. Then $Y(r)$ \dgy{varies monotonically  
from $Y_0$ at  $r=R_2$ to
$(1+\dy{w})\,Y_0$ at $r=R_1$}.
We have tried some other functions which 
amplified variations near the surface or near
the crust bottom, but they have not led to
much better variational
estimates of $\nu$. Since our approach is oversimplified
anyway (e.g., it 
neglects possible mixtures of states with
different $\ell$ and $m$), we would not like 
to complicate the variational function. 

Figure \ref{f:freqvar} compares the FOPT
frequencies, calculated at $\ell=4$ and
$\ell=7$ (black curves) with corresponding
variational estimates (blue curves) for $B_0 \leq 4 \times
10^{15}$ G. Variational estimates are 
naturally smaller than the FOPT values. As expected, the difference grows up with 
increasing $B_0$ and $m$. At $\ell=7$
the difference is more pronounced, than at
$\ell=4$, and becomes visible at lower $B_0$.
Otherwise, both approaches seem to be in a reasonable qualitative agreement. This allows us to
expect that using FOPT at $B_0 \lesssim 4 \times 10^{15}$ G does not lead to serious errors.  

\subsection{Zeeman splitting of fine structures of ordinary modes
($n>0$)}
\label{s:freqn1}

Let us focus on ordinary magneto-elastic modes which exist in
addition to fundamental modes and which are characterized by
the presence of nodes of the radial wave function $Y(r)$. 
Typical frequencies of these modes are higher than several
hundred Hz. We restrict ourselves by the FOPT approximation. 

The ordinary modes are different from fundamental ones.
In the absence of a magnetic field, for a fixed $n>0$ there
exist a family of closely spaced 
frequencies $\nu_{\mu n \ell}$ which slowly
grow up with increasing $\ell=2,3,\ldots$ Their
spectrum resembles fine structure of atomic
energy levels. Such spectra have been studied
in Ref.\ \cite{2023Yak2} in full GR, with the 
result that for any fixed $n$ they
are accurately fitted as
\begin{equation}
    \nu^2_{\mu n\ell}= \nu^2_{\mu n}+ 
    (\ell+2)(\ell-1) \delta \nu^2_{\mu n},
\label{e:numun}     
\end{equation} 
with only two constant auxiliary frequencies,
$\nu_{\mu n}$ and $\delta \nu_{\mu n}$. 

The magnetic correction $\nu_{B n \ell m}$ can
be calculated using Eq.\ (\ref{e:omegaB6}) with
the wave functions $Y_{n\ell}(r)$ computed in the same
way as described in Ref.\ \cite{2023Yak1}. The
result can be written in the
form similar to Eq.\ (\ref{e:omegaB7a}),
\begin{equation}
\nu_{Bn\ell m}^2=m^2 \nu_{Bn_*}^2 B_*^2,
\label{e:omegaB7b}	
\end{equation}
where $\nu_{Bn_*}$ is a single `magnetic' auxiliary frequency 
for a fixed $n$. It is independent of $\ell$ 
because $Y_{n\ell}(r)$ is nearly independent of $\ell$
at $n>0$ \cite{2023Yak2}. 

\begin{figure}[t]
	\includegraphics[width=0.5\textwidth]{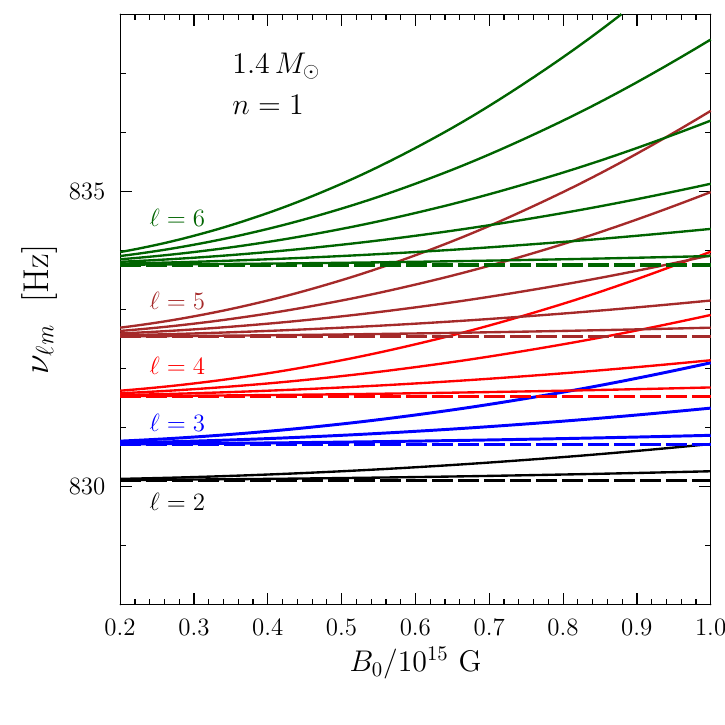}
	\caption{\label{f:n1lowB} Zeeman
		splitting versus $B_0$ of fine structure components of oscillations with one radial node. We plot the splitting for bunches with $\ell$ from 2 to 6. Adding
		bunches with larger $\ell$ would 
		fill the range of higher 
		$\nu$. For example,
        at $\ell=$7, 8, 9 and 10, the basic fine-structure
        frequencies (\ref{e:numun}) $\nu_{\mu 1\ell}$ are 835.2,
        836.8, 8.38.6 and 840.6, respectively.
	}
\end{figure}

\begin{figure}[th]
	\includegraphics[width=0.48\textwidth]{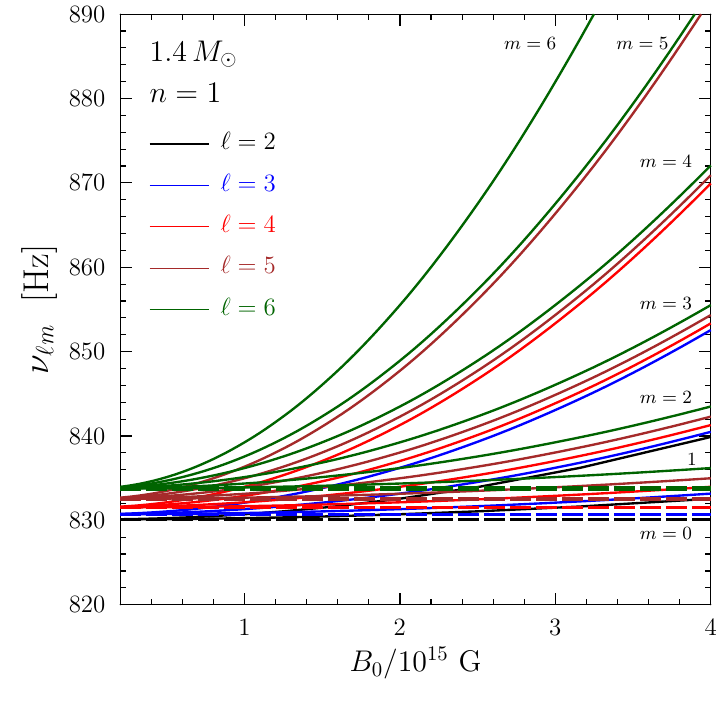}
	\caption{\label{f:n1} Same as in Fig.\ \ref{f:n1lowB} but for a wider
	range of $B_0$ and $\nu$.}
\end{figure}
 
Now a squared total FOPT frequency of an ordinary
magneto-elastic oscillation is
\begin{equation}
\nu^2_{nlm}=\nu^2_{\mu n}+
(\ell+2)(\ell-1) \delta \nu^2_{\mu n}+m^2 \nu_{Bn_*}^2 B_*^2.
\label{e:freqn1}
\end{equation}  
It is determined by the three constants, $\nu_{\mu n}$,
$\delta \nu_{\mu n}$ and $\nu^2_{B n*}$ for
a given fine structure with fixed $n$.

For our $1.4\,\msun$ star, these
constants, with $n=1$ and 2, are listed in Table \ref{tab:main1}: $\nu_{\mu n}$ and 
$\delta \nu_{\mu n}$ are taken from \cite{2023Yak2},
while the constants $\nu_{B n*}$ are original,
calculated assuming $\psi(x)=\psi_0(x)$.

For example, Figs.\ \ref{f:n1lowB} and \ref{f:n1}
show the fine structure and its Zeeman splitting 
for a family of oscillation frequencies with
$n=1$, for values of $\ell$ from 2 to 6, and
all values of $m$. Bunches of curves with higher
$\ell=7,8,\dots$ are not plotted for simplicity.  
Figure \ref{f:n1lowB} is restricted by
$B_0 \leq 10^{15}$ G and by a narrow frequency 
interval of the width $\approx$ 10 Hz.
Figure \ref{f:n1} is an extension to $B_0 \leq 4 \times 10^{15}$ 
and to higher $\nu$ for demonstrating a general
view. 
 
The fine structure itself ($B_0=0$) is shown
by horizontal dashed lines. The lowest frequency is
830.1 Hz. All five plotted frequencies fall within the
4 Hz window. 

The Zeeman effect splits the fine-structure components
into the bunches of modes with the same $\ell$ but different $m$. Because the difference of frequencies between fine-structure
components is small, the bunches start to cross (Fig.\ \ref{f:n1lowB}) at smaller
$B_0 \lesssim 10^{15}$ G than for the fundamental modes
(Sec.\ \ref{s:n014msun}). 

At $B_0 \gtrsim
10^{15}$ G one gets distinctly different 
bunches (Fig.\ \ref{f:n1}). These can
can be numbered by the index $m$, and the curves
within any buch can be numbered by $\ell$. Such rearrangement
of the structure of bunches resembles the well known Paschen-Back effect in atomic physics
(e.g., \cite{LL76}). The higher $\ell$, the lower $B_0$ 
at which the rearrangement occurs.

Adding oscillation frequencies with higher $\ell$, we
would obtain a dense sequence of magneto-elastic oscillations at the same $n=1$. Any high-frequency magnetar QPO observed in this range would be consistent
with this theory. This adds to our discussion 
(Sec.\ \ref{s:n014msun}) of allowed (darkened) regions of oscillation frequencies
in Fig.\ \ref{f:freqn0a}.

The situation with cases of two and more 
nodes ($n\geq 2$) is similar. The frequencies
of these oscillations are even higher. For instance,
the lowest frequency for the same neutron star model
at $n=2$ is 1.327 kHz.

\subsection{Self-similarity. Different masses}
\label{s:selfsim}

\renewcommand{\arraystretch}{1.2}
\begin{table*}[ht]
	\caption{\label{tab:main2}%
		Fit parameters in
		Eqs.\ (\ref{e:Fitdnuln}), (\ref{e:Fitnuln}) 
		and  (\ref{e:FitdnulnB}) which determine magneto-elastic oscillations 
		of fundamental modes ($n=0$) and ordinary modes with $n=1$ and $n=2$ 
		for neutron stars models with the BSk21 EOS at $M \geq 1 \ \msun $ }
	\begin{ruledtabular}	
		\begin{tabular}{c c c c c c c c c c}
			$n$ & 
			$\delta f_{\mu n}$ [Hz] &  $\delta \alpha_{\mu n}$ & $\delta \beta_{\mu n}$ &
			$ f_{\mu n}$ [Hz] &   $\alpha_{\mu n}$ & $ \beta_{\mu n}$ & 
			$\delta f_{B n*}$ [Hz] &  $\delta \alpha_{B n*}$ 
			& $ \delta \beta_{B n*}$\\ 
			\colrule
			0	&	44.59 &  2.411 &  $-$1.968 &	0 &  -- &  --  &   9.940 &  5.662 &  $-$4.504
			\\
			1	&	24.61 &  2.166 & --1.801 & 1171.7 &  $-$1.508 & 1.326 
			& 40.00  & 6.779 & $-$ 5.381 \\
			2	&	27.35 &  1.750 & --1.340 &	1821.5 & $-$1.342 & 1.163 &
			61.01 & 8.000 & $-$5.923
			\\
		\end{tabular}
	\end{ruledtabular}	
\end{table*}
\renewcommand{\arraystretch}{1.0} 
%
%
%
%
%

Table \ref{tab:main1} contains eight constants which
are sufficient to calculate frequencies of
magneto-elastic oscillations with $n=$0, 1 and 2, for
our $1.4 \msun$ neutron star model with the toroidal
magnetic field geometry defined by Eq.\ (\ref{e:psi0}). 

Now we employ the self-similarity relations \cite{2023Yak2}, which allow one to extend
these results to a range of stellar masses with fixed 
EOS. For instance, in Ref.\ \cite{2023Yak2}, 
the auxiliary frequencies denoted here as $\nu_{\mu 0}$,
as well as $\nu_{\mu n}$ and  $\delta \nu_{\mu n}$ 
with $n>0$, were calculated at $M/\msun$=1, 1.2, 1.4,
\ldots 2.2 (the BSk21 EOS) 
and fitted by simple expressions.
To simplify using these results, all 
frequencies were fitted \cite{2023Yak2} by analytic functions of
neutron star mass and radius, 
\begin{equation}
\delta {\nu}_{\mu n}
= 
\frac{\sqrt{1-\xg } \, \delta f_{\mu n}}{R_{10} 
	\sqrt{1+ \delta \alpha_{\mu n} \xg
		+ \delta \beta_{\mu n} x_{ g}^2}},
\label{e:Fitdnuln}	
\end{equation}
\begin{equation}
\nu_{\mu n }
=\frac{M_1}{R_{10}^2} f_{\mu n}
\sqrt{1+\alpha_{\mu n}  \xg + \beta_{\mu n} x_{ g}^2},
\label{e:Fitnuln}
\end{equation}
where  $\xg$ is defined in Eq.\ (\ref{e:toymetric}); $R_{10}=R/10$ km,
$M_1= M/\msun$; $f_{\mu n}$, $\alpha_{\mu n}$, $\beta_{\mu n}$,  $\delta f_{\mu n}$,
$\delta  \alpha_{\mu n}$ and $\delta \beta_{\mu n}$ are fit parameters.
For $n$=0, 1 and 2 we
reproduce these fit parameters in Table \ref{tab:main2}. 
Note that 
$\nu_{\mu 0}$, $\delta \nu_{\mu 1}$ and $\delta \nu_{\mu 2}$
have to be calculated using Eq.\ (\ref{e:Fitdnuln}),
while  $\nu_{\mu 1}$ and $\nu_{\mu 2}$ -- using Eq.\ (\ref{e:Fitnuln}).

Here we compute the magnetic auxiliary frequencies $\nu_{nB_*}$ 
at $n$=0, 1 and 2 on the same grid of masses. Considering self-similarity of such frequencies in the same manner
as in \cite{2023Yak2}, we obtain that they can be
fitted by 
\begin{equation}
\delta {\nu}_{Bn*}
= 
\frac{\sqrt{1-\xg } \, \delta f_{B n}}{R_{10} 
	\sqrt{1+ \delta \alpha_{B n} \xg
		+ \delta \beta_{B n} \xg^2}},
\label{e:FitdnulnB}	
\end{equation}
where $\delta f_{Bn}$,
$\delta  \alpha_{Bn}$ and $\delta \beta_{Bn}$ 
are the 
fit parameters which we list in Table \ref{tab:main2}. 

As a result, Table \ref{tab:main2} contains 24 fit 
parameters which allows one to calculate all FOPT
frequencies of magneto-elastic oscillations for
fundamental modes ($n=0$), and
two series of ordinary modes ($n=$1 and 2) in
neutron stars with the BSk21 EOS at any mass $M$ ranged
from $1\, \msun$ to $2.2\, \msun$. These stars possess
toroidal magnetic field localized
in their crust and  obeying Eq.\ (\ref{e:psi0}).

Note that in the indicated mass range the radii $R$ (in km)
of these stars can be fitted  by
\begin{equation}
R=R_{\rm min}+3.552\, (M/M_{\rm max}) \sqrt{1-(M/M_{\rm max})}, 
\label{e:Rfit}
\end{equation}
where $R_{\rm min}=11.246$ km is the radius of the star with
the maximum mass $M_{\rm max}=2.27 \, \msun$. 
The maximum relative fit error is 0.6 percent.
This facilitates
calculations of oscillation frequencies.

%
%
%
%

The suggested analytic fits are fairly accurate: the fitted  
oscillation frequencies do not deviate from the initially calculated ones by more than a few percent. This is much better than needed for analyzing current data on magnetar QPOs (Sec.\ \ref{s:discuss}).
Besides, one cannot seriously believe that magnetars possess
purely toroidal magnetic fields of this specific geometry, and that
the FOPT is the best tool to describe the oscillations.  

Moreover, we had no intention to obtain so accurate fits:
they seem to result from natural self-similarity relations
\cite{2023Yak2}. As shown in \cite{2023Yak2},
similar relations are
valid for a wide class of EOSs. We do not consider other
EOSs here because we do not expect to obtain anything qualitatively different.

One can formulate other self-similarity relations which
will be discussed in the next section.  

\section{Discussion}
\label{s:discuss}

\begin{figure}[b]
	\includegraphics[width=0.5\textwidth]{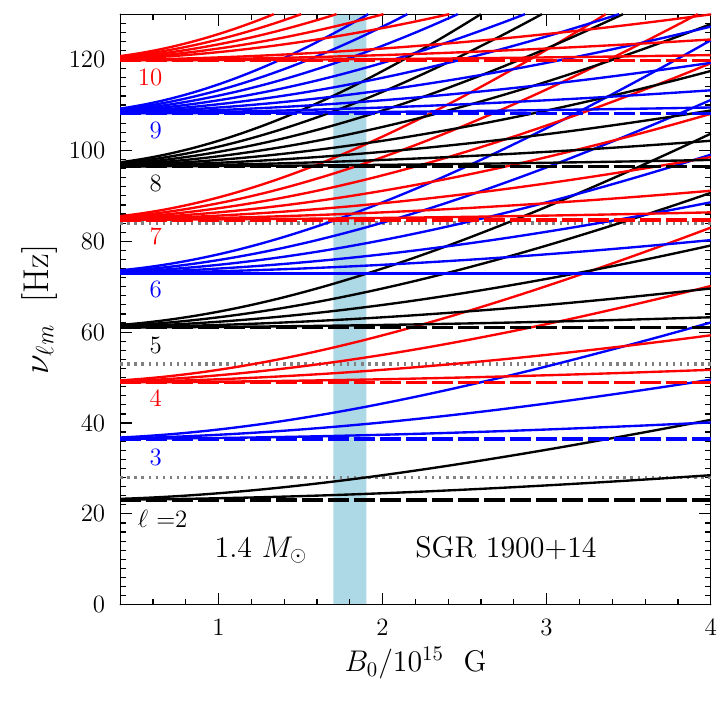}
	\caption{\label{f:1900} Calculated frequencies of magneto-elastic oscillations versus
	$B_0$ for a $1.4\, \msun $ magnetar model with the BSk21 EOS (from Fig.\ \ref{f:freqn0}) are confronted
    with observations of QPOs from the
    giant flare of SGR 1900+14 (dotted
    lines). For illustration, the vertical darkened strip shows a possible range of $B_0$ for this
    particular toy model. See the text
    for details. }
\end{figure}

\begin{figure}[b]
	\includegraphics[width=0.5\textwidth]{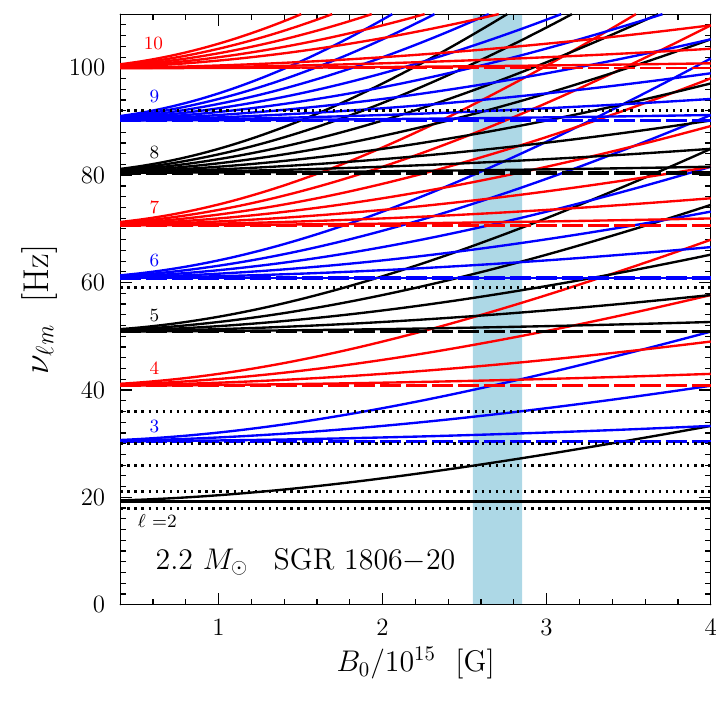}
	\caption{\label{f:1806} Same as in Fig.\ \ref{f:1900}
	but employing the $2.2 \msun$ neutron star model for explaining QPOs observed from the
    hyperflare of SGR 1806--20.}
\end{figure}

We have calculated frequencies of magneto-elastic oscillations in the crust of neutron stars with toroidal
magnetic fields of specific geometry, (\ref{e:B_phi}) and
(\ref{e:psi0}), fully localized in the crust. This allows us to avoid studying 
leakage of Alfv\'en perturbations outside the crust (Sec.\
\ref{s:alfven}) and makes the theory more reliable. The assumed $\B(\bm{r})$ structure is
just a convenient simplified toy model, which can reflect general features of magneto-elastic oscillations. 

The QPO data were mostly obtained
from observations of the giant flare from the SGR 1900+14 and the hyperflare from the SGR 1806--20.
The QPO frequencies were found by processing and reprocessing
the observed ligtcurves of these flaring magnetars,  
e.g., Refs.\
\cite{2005Israel,2006Watts,2011Hambaryan,2014Huppen,2014Huppenkothen}. The results are summarized, for instance, in Ref.\ \cite{2018Gabler}.  The QPOs at frequencies 
28, 53,
84 and 155 Hz were detected from the giant flare of
SGR 1900+14. The QPOs at  
18, 26, 30, 92 and 150 Hz (as well as at 17, 21, 36, 59 and 116 Hz
with a lower significance) were found from the hyperflare
of SGR 1806--20.

Nevertheless, there was one publication, Ref.\ \cite{2018Pumpe}, based on a procedure of 
removing noise of observed signals at $\nu \lesssim 150$ Hz using \dy{some original version of} Bayesian
analysis. It did not confirm statistical significance 
of all low-frequency QPOs reported previously from both flares but discovered two new 
QPOs at 9.2 and 7.7 Hz from the SGR 1806--20 hyperflare. An additional \dy{confirmation} of these results, reported by one group in one
publication, would be 
welcome but it has not appeared in six years. In its absence, we follow Refs.\ \cite{2018Gabler,2018Sotani} and employ the familiar set of observable QPOs which was accepted 
and analyzed in many publications. 

Although strict explanation of the data with our toy 
model cannot be serious, Figs.\ \ref{f:1900} and \ref{f:1806} present two illustrative possibilities. 

In Fig.\ \ref{f:1900} we compare the frequencies of three QPOs
(dotted lines),
supposed to be observed from the giant flare of SGR 1900+14,
with the frequencies of magneto-elastic
oscillations calculated (Fig.\ \ref{f:freqn0})
for our 1.4 $\msun$ magnetar model at different
values of $B_0$. 
By choosing $B_0=(1.7 - 1.9) \times 10^{15}$ G, we can
explain these QPOs assuming excitation of oscillations with
certain $\ell$ and $m$. The QPO at 28 Hz corresponds to 
$\ell=2$ and $m=2$; the QPO at 53 Hz -- to $\ell=4$ and
$m=3$; while the 84 Hz QPO corresponds  either to $\ell=m=6$ or to 
$\ell=7$ and $m<2$. The forth QPO at 155 Hz is not plotted but it is evidently explained in this model.    
   
Figure \ref{f:1806} is similar but designed to 
interpret much richer spectrum of QPOs observed from the hyperflare of SGR 1806--20. These
QPOs cannot be explained assuming the $1.4 \msun$ neutron star. A higher mass is required for the QPOs with
lowest frequencies (e.g. Refs.\ \cite{2023Yak1,2024Yak}): higher $M$
reduces oscillation frequencies due to stronger
gravitational redshift. In Fig.\ \ref{f:1806} we assume the $2.2 \msun$ star.
Choosing $B_0$ in the range $(2.7 \pm 0.12) \times 10^{15}$ G, we can be reasonably consistent with the data except
for the triplet of QPOs
with lowest frequencies (17, 18 and 21 Hz). 
It is
not explained by our particular model, but 
can be easily attributed to deviations from
the idealized $\B(\bm{r})$ configuration, given
by Eqs.\ (\ref{e:B_phi}) and
(\ref{e:psi0}). 

These exercises demonstrate once more that including 
\dy{the} Zeeman splitting
greatly \dy{affects} interpretation of observed QPOs. In fact, it is sufficient to focus on explaining low frequency
oscillations ($\nu \lesssim 100$ Hz), because the region
of higher frequencies will be densely covered by Zeeman structures anyway. All low-frequency oscillations belong
to fundamental modes which consideration is especially
simple.

The model of toroidal magnetic fields studied above
is the simplest one. Because of that we can formulate additional
self-similarity rules, specific for this particular model.  
One needs only two numbers, $\nu_{\mu 0}$ and $\nu_{B0*}$, to construct the FOPT oscillation frequencies, given by Eq.\ (\ref{e:nu0}). To interpret the QPOs from one flare of a given SGR,
it would be highly unreasonable to fix the neutron star model. 
It would be more profitable to consider
$\nu_{\mu 0}$ and  the product $(\nu_{B0*} B_*)$ as two independent fit
parameters and try to find reliable best fits. In case
of success, at the next step one can analyse these two best fit values with the hope to 
understand which neutron star and magnetic field parameters are physically sound. Because the fit gives the product
$(\nu_{B0*} B_*)$, there appears internal ambiguity of the interpretation: one and the same value of the product will correspond to a family of different magnetic strengths $B_0$ and geometries, as well as different parameters of magnetar crust. 

Let us remark that in Ref.\ \cite{2023Yak1,2024Yak} we have studied 
similar magneto-elastic oscillations of magnetars using FOPT but
assuming purely dipole magnetic field in the crust. Those results 
can be sensitive to a leakage of Alfv\'en perturbations outside the crust that was neglected. Nevertheless, the calculated
spectra of magneto-elastic oscillations with account of \dy{the} Zeeman
splitting appeared rather similar 
to those obtained here (the largest difference was that the magnetic frequency $\nu_{B}$ depended on $\ell$ and 
did not vanish at $m=0$). 
Illustrative attempts \cite{2023Yak1,2024Yak} to interpret the same spectra of QPOs in the flaring SGR 1900+14 and 
SGR 1860--20 led to the same conclusions as here: 
(i) the SGR 1900+14 may have
\dy{a typical neutron-star} mass but the SGR 1860--20 needs 
to be massive; (ii)
one needs magnetic fields $B \sim$  a few times of $10^{15}$ G in
both SGRs, and the field of SGR 1806--20 is somewhat higher. 

\section{Conclusions}
\label{s:conclude}

We have studied \dy{the} Zeeman splitting of magneto-elastic oscillation 
frequencies of soft-gamma repeaters (SGRs), which are magnetars --
neutron stars possessing very strong magnetic fields. 
\dy{ The QPOs appear 
in occasional powerful flares of SGRs and can contain potentially important information
on the physics of magnetar structure and evolution}.

\dy{The  Zeeman splitting of magnetar QPOs was predicted
in Ref.\ \cite{2009SE}. Afterwards its study has been elaborated in
two recent publications}
\cite{2023Yak1,2024Yak}. 

In Sec.\ \ref{s:theory} we \dy{present} the first-order perturbation
theory (FOPT) of magneto-elastic oscillations localized
in the magnetar crust, assuming axial symmetry of the crustal
magnetic field $\B(\bm{r})$ and the presence of both, poloidal and
toroidal, field components. 

In Sec.\ \ref{s:toroidal} we \dy{apply} these results for studying magneto-elastic oscillations in the presence of purely toroidal
magnetic field, which forbids propagation of Alfv\'en perturbations
outside the crust. The toroidal field \dy{is} taken in the form (\ref{e:B_phi}). We \dy{describe} amazing simplicity of the spectrum of fundamental oscillations (without radial nodes) and checked the 
validity of FOPT calculations using a restricted variational method. We \dy{analyse} ranges of frequencies densely covered by numerous Zeeman components; at higher frequencies these ranges occur in 
lower magnetic fields. These results are in good qualitative agreement with those obtained previously \cite{2023Yak1,2024Yak}
for dipole magnetic fields in the magnetar crust. In addition, we
\dy{study}, for the first time, the effects of magnetic
fields on subsets of fine-structure components
of ordinary oscillations (with radial nodes). Their
Zeeman spitting creates additional frequency
ranges densely filled with Zeeman structures.

Our calculations \dy{are} mostly performed for the $1.4 \msun$ magnetar
model with the BSk21 EOS (Table \ref{tab:main1}) but the self-similarity relations \cite{2023Yak2}
\dy{allow} us to extend the results to wider range of neutron star masses and describe oscillation spectra
of numerous modes by a very limited amount of constants
(Tables \ref{tab:main1} and \ref{tab:main2}). Similar
description can done for other EOSs.
 
Our schematic analysis of QPOs observed during the giant
flare of SGR 1900+14 and the hyperflare of SGR 1806--20 supports previous conclusions \cite{2009SE,2023Yak1,2024Yak} that the Zeeman effect is 
\dy{very} important for a proper interpretation of the
data. The main focus should be on low-frequency QPOs
($\nu \lesssim 150$ Hz) which can be interpreted as
produced by fundamental magneto-acoustic oscillations
at not too high magnetic fields $B \lesssim 3 \times 10^{15}$ G. Higher frequency QPOs  
fall in the range of frequencies densely covered 
by Zeeman structures even at lower $B$. The presented theory can identify them but the identification will be
highly ambiguous.

On the other hand, we should state that neither theory
not observations are ready for the detailed analysis.
The status of observations is outlined in
Sec.\ \ref{s:discuss}. It has to be 
clarified. The
present version of the theory can be improved in many
ways. It would be more realistic to consider the 
magnetar's magnetic field as a combination of 
toroidal and poloidal components (Sec.\ \ref{s:theory}).
In such studies, one needs to go beyond the FOPT,
analyse quasi-crossings of different
modes and propagation of Alfv\'en perturbations outside
the neutron star crust \dy{(Sec.\ \ref{s:alfven})}. It is also desirable to solve the whole problem in full GR.

\dy{Let us stress again that the 
present calculations are performed for one EOS of
neutron star matter and standard microphysics of the crust. One can
vary the EOS and crustal microphysics, including different models of shear modulus, possible nuclear pasta layers at the crust bottom and nucleon superfludity there, as discussed, for instance, in Sec.\ \ref{s:discuss} and in Refs.\
\cite{2023Yak1,2023Yak2}. This will affect the interpretation of QPOs at the stage of accurate quantitative analysis.}

\dy{Combining
our results with the formalism 
of solving the Alfv'en wave problem (Sec.\ \ref{s:alfven})
seems most important. Assuming
a purely toroidal magnetic field localized in the crust, we artificially
avoided it in this paper, but the step is inevitable. In more realistic cases, Alfv\'enic
perturbations propagate into the stellar core, interact with
the Alfv\'enic perturbations available there. This can affect crust-core coupling, damping of
magneto-elastic oscillations and loss of their coherence. As a result, magnetar QPOs will depend on the structure,
composition and microphysics of the entire star (particularly, on
nucleon superlfluidity and magnetic field geometry in the core)
in a complicated way. As mentioned in Sec.\ \ref{s:alfven}, this effect has been studied in
numerous publications. Unfortunately, all of them have been restricted 
by axially symmetric perturbations, while we stress
the importance of axially asymmetric perturbations and 
suggest to elaborate the Alfv\'enic wave problem accordingly. Such
elaboration would require a lot of efforts.
}

\dy{In addition, let us mention the long-standing problem of 
QPO formation, meaning detailed mechanism for transforming oscillations
of surface neutron-star layers into observable QPOs. The problem seems important and feasible but remains unsolved.}

\dy{Finally, we note that by} demonstrating Zeeman splitting of oscillations on
macroscopic (stellar) level, a magnetar behaves like
a microscopic quantum object, resembling one
giant atomic nucleus, just as Lev Landau envisioned 
a neutron star \cite{1932Landau} 
in the very beginning of 1930s (see Ref.\ \cite{2013Yak} for details). 

\begin{acknowledgments}
\dy{We are indebted to the anonymous referee for many useful
remarks and critical comments.}
DY is grateful to M.~E.\ Gusakov and E.~M.\ Kantor
for encouraging to study toroidal magnetic fields
localized in the crust.	
The work was supported by the Russian Science
Foundation, grant number 24-12-00320. 
\end{acknowledgments}

\newcommand{\araa}{Ann. Rev. Astron. Astrophys.}
\newcommand{\aap}{Astron. Astrophys.}
\newcommand{\aj}{Astron. J.}
\newcommand{\apjl}{Astrophys. J. Lett.}
\newcommand{\apjs}{Astrophys. J. Suppl. Ser.}
\newcommand{\apss}{Astrophys. Space Sci.}
\newcommand{\mnras}{Mon. Not. R. Astron. Soc.}
\newcommand{\pasa}{Publ. Astron. Soc. Aust.}
\newcommand{\pasj}{Publ. Astron. Soc. Jpn.}
\newcommand{\pasp}{Publ. Astron. Soc. Pac.}
\newcommand{\qjras}{Q. J. R. Astron. Soc.}
\newcommand{\sovast}{Sov. Astron.}
\newcommand{\ssr}{Space Sci. Rev.}



\begin{thebibliography}{57}%
	\makeatletter
	\providecommand \@ifxundefined [1]{%
		\@ifx{#1\undefined}
	}%
	\providecommand \@ifnum [1]{%
		\ifnum #1\expandafter \@firstoftwo
		\else \expandafter \@secondoftwo
		\fi
	}%
	\providecommand \@ifx [1]{%
		\ifx #1\expandafter \@firstoftwo
		\else \expandafter \@secondoftwo
		\fi
	}%
	\providecommand \natexlab [1]{#1}%
	\providecommand \enquote  [1]{``#1''}%
	\providecommand \bibnamefont  [1]{#1}%
	\providecommand \bibfnamefont [1]{#1}%
	\providecommand \citenamefont [1]{#1}%
	\providecommand \href@noop [0]{\@secondoftwo}%
	\providecommand \href [0]{\begingroup \@sanitize@url \@href}%
	\providecommand \@href[1]{\@@startlink{#1}\@@href}%
	\providecommand \@@href[1]{\endgroup#1\@@endlink}%
	\providecommand \@sanitize@url [0]{\catcode `\\12\catcode `\$12\catcode
		`\&12\catcode `\#12\catcode `\^12\catcode `\_12\catcode `\%12\relax}%
	\providecommand \@@startlink[1]{}%
	\providecommand \@@endlink[0]{}%
	\providecommand \url  [0]{\begingroup\@sanitize@url \@url }%
	\providecommand \@url [1]{\endgroup\@href {#1}{\urlprefix }}%
	\providecommand \urlprefix  [0]{URL }%
	\providecommand \Eprint [0]{\href }%
	\providecommand \doibase [0]{https://doi.org/}%
	\providecommand \selectlanguage [0]{\@gobble}%
	\providecommand \bibinfo  [0]{\@secondoftwo}%
	\providecommand \bibfield  [0]{\@secondoftwo}%
	\providecommand \translation [1]{[#1]}%
	\providecommand \BibitemOpen [0]{}%
	\providecommand \bibitemStop [0]{}%
	\providecommand \bibitemNoStop [0]{.\EOS\space}%
	\providecommand \EOS [0]{\spacefactor3000\relax}%
	\providecommand \BibitemShut  [1]{\csname bibitem#1\endcsname}%
	\let\auto@bib@innerbib\@empty
	\bibitem [{\citenamefont {{Shapiro}}\ and\ \citenamefont
		{{Teukolsky}}(1983)}]{ST1983}%
	\BibitemOpen
	\bibfield  {author} {\bibinfo {author} {\bibfnamefont {S.~L.}\ \bibnamefont
			{{Shapiro}}}\ and\ \bibinfo {author} {\bibfnamefont {S.~A.}\ \bibnamefont
			{{Teukolsky}}},\ }\href@noop {} {\emph {\bibinfo {title} {{Black holes, white
					dwarfs, and neutron stars: The physics of compact objects}}}}\ (\bibinfo
	{publisher} {Wiley-Interscience, New York},\ \bibinfo {year}
	{1983})\BibitemShut {NoStop}%
	\bibitem [{\citenamefont {{Haensel}}\ \emph {et~al.}(2007)\citenamefont
		{{Haensel}}, \citenamefont {{Potekhin}},\ and\ \citenamefont
		{{Yakovlev}}}]{HPY2007}%
	\BibitemOpen
	\bibfield  {author} {\bibinfo {author} {\bibfnamefont {P.}~\bibnamefont
			{{Haensel}}}, \bibinfo {author} {\bibfnamefont {A.~Y.}\ \bibnamefont
			{{Potekhin}}},\ and\ \bibinfo {author} {\bibfnamefont {D.~G.}\ \bibnamefont
			{{Yakovlev}}},\ }\href@noop {} {\emph {\bibinfo {title} {{Neutron Stars. 1.
					Equation of State and Structure}}}}\ (\bibinfo  {publisher} {Springer, New
		York},\ \bibinfo {year} {2007})\BibitemShut {NoStop}%
	\bibitem [{\citenamefont {{Kaspi}}\ and\ \citenamefont
		{{Beloborodov}}(2017)}]{2017KB}%
	\BibitemOpen
	\bibfield  {author} {\bibinfo {author} {\bibfnamefont {V.~M.}\ \bibnamefont
			{{Kaspi}}}\ and\ \bibinfo {author} {\bibfnamefont {A.~M.}\ \bibnamefont
			{{Beloborodov}}},\ }\bibfield  {title} {\bibinfo {title} {{Magnetars}},\
	}\href@noop {} {\bibfield  {journal} {\bibinfo  {journal} {\araa}\ }\textbf
		{\bibinfo {volume} {55}},\ \bibinfo {pages} {261} (\bibinfo {year} {2017})},\
	\Eprint {https://arxiv.org/abs/1703.00068} {arXiv:1703.00068 [astro-ph]}
	\BibitemShut {NoStop}%
	\bibitem [{\citenamefont {{Duncan}}(1998)}]{1998Duncan}%
	\BibitemOpen
	\bibfield  {author} {\bibinfo {author} {\bibfnamefont {R.~C.}\ \bibnamefont
			{{Duncan}}},\ }\bibfield  {title} {\bibinfo {title} {{Global Seismic
				Oscillations in Soft Gamma Repeaters}},\ }\href
	{https://doi.org/10.1086/311303} {\bibfield  {journal} {\bibinfo  {journal}
			{\apjl}\ }\textbf {\bibinfo {volume} {498}},\ \bibinfo {pages} {L45}
		(\bibinfo {year} {1998})},\ \Eprint {https://arxiv.org/abs/astro-ph/9803060}
	{arXiv:astro-ph/9803060 [astro-ph]} \BibitemShut {NoStop}%
	\bibitem [{\citenamefont {{Israel}}\ \emph {et~al.}(2005)\citenamefont
		{{Israel}}, \citenamefont {{Belloni}}, \citenamefont {{Stella}},
		\citenamefont {{Rephaeli}}, \citenamefont {{Gruber}}, \citenamefont
		{{Casella}}, \citenamefont {{Dall'Osso}}, \citenamefont {{Rea}},
		\citenamefont {{Persic}},\ and\ \citenamefont {{Rothschild}}}]{2005Israel}%
	\BibitemOpen
	\bibfield  {author} {\bibinfo {author} {\bibfnamefont {G.~L.}\ \bibnamefont
			{{Israel}}}, \bibinfo {author} {\bibfnamefont {T.}~\bibnamefont {{Belloni}}},
		\bibinfo {author} {\bibfnamefont {L.}~\bibnamefont {{Stella}}}, \bibinfo
		{author} {\bibfnamefont {Y.}~\bibnamefont {{Rephaeli}}}, \bibinfo {author}
		{\bibfnamefont {D.~E.}\ \bibnamefont {{Gruber}}}, \bibinfo {author}
		{\bibfnamefont {P.}~\bibnamefont {{Casella}}}, \bibinfo {author}
		{\bibfnamefont {S.}~\bibnamefont {{Dall'Osso}}}, \bibinfo {author}
		{\bibfnamefont {N.}~\bibnamefont {{Rea}}}, \bibinfo {author} {\bibfnamefont
			{M.}~\bibnamefont {{Persic}}},\ and\ \bibinfo {author} {\bibfnamefont
			{R.~E.}\ \bibnamefont {{Rothschild}}},\ }\bibfield  {title} {\bibinfo {title}
		{{The Discovery of Rapid X-Ray Oscillations in the Tail of the SGR 1806-20
				Hyperflare}},\ }\href {https://doi.org/10.1086/432615} {\bibfield  {journal}
		{\bibinfo  {journal} {\apjl}\ }\textbf {\bibinfo {volume} {628}},\ \bibinfo
		{pages} {L53} (\bibinfo {year} {2005})},\ \Eprint
	{https://arxiv.org/abs/astro-ph/0505255} {arXiv:astro-ph/0505255 [astro-ph]}
	\BibitemShut {NoStop}%
	\bibitem [{\citenamefont {{Strohmayer}}\ and\ \citenamefont
		{{Watts}}(2005)}]{2005Strohmayer}%
	\BibitemOpen
	\bibfield  {author} {\bibinfo {author} {\bibfnamefont {T.~E.}\ \bibnamefont
			{{Strohmayer}}}\ and\ \bibinfo {author} {\bibfnamefont {A.~L.}\ \bibnamefont
			{{Watts}}},\ }\bibfield  {title} {\bibinfo {title} {{Discovery of Fast X-Ray
				Oscillations during the 1998 Giant Flare from SGR 1900+14}},\ }\href
	{https://doi.org/10.1086/497911} {\bibfield  {journal} {\bibinfo  {journal}
			{\apjl}\ }\textbf {\bibinfo {volume} {632}},\ \bibinfo {pages} {L111}
		(\bibinfo {year} {2005})},\ \Eprint {https://arxiv.org/abs/astro-ph/0508206}
	{arXiv:astro-ph/0508206 [astro-ph]} \BibitemShut {NoStop}%
	\bibitem [{\citenamefont {{Watts}}\ and\ \citenamefont
		{{Strohmayer}}(2006)}]{2006Watts}%
	\BibitemOpen
	\bibfield  {author} {\bibinfo {author} {\bibfnamefont {A.~L.}\ \bibnamefont
			{{Watts}}}\ and\ \bibinfo {author} {\bibfnamefont {T.~E.}\ \bibnamefont
			{{Strohmayer}}},\ }\bibfield  {title} {\bibinfo {title} {{Detection with
				RHESSI of High-Frequency X-Ray Oscillations in the Tailof the 2004 Hyperflare
				from SGR 1806-20}},\ }\href {https://doi.org/10.1086/500735} {\bibfield
		{journal} {\bibinfo  {journal} {\apjl}\ }\textbf {\bibinfo {volume} {637}},\
		\bibinfo {pages} {L117} (\bibinfo {year} {2006})},\ \Eprint
	{https://arxiv.org/abs/astro-ph/0512630} {arXiv:astro-ph/0512630 [astro-ph]}
	\BibitemShut {NoStop}%
	\bibitem [{\citenamefont {{Hambaryan}}\ \emph {et~al.}(2011)\citenamefont
		{{Hambaryan}}, \citenamefont {{Neuh{\"a}user}},\ and\ \citenamefont
		{{Kokkotas}}}]{2011Hambaryan}%
	\BibitemOpen
	\bibfield  {author} {\bibinfo {author} {\bibfnamefont {V.}~\bibnamefont
			{{Hambaryan}}}, \bibinfo {author} {\bibfnamefont {R.}~\bibnamefont
			{{Neuh{\"a}user}}},\ and\ \bibinfo {author} {\bibfnamefont {K.~D.}\
			\bibnamefont {{Kokkotas}}},\ }\bibfield  {title} {\bibinfo {title} {{Bayesian
				timing analysis of giant flare of SGR 180620 by RXTE PCA}},\ }\href
	{https://doi.org/10.1051/0004-6361/201015273} {\bibfield  {journal} {\bibinfo
			{journal} {\aap}\ }\textbf {\bibinfo {volume} {528}},\ \bibinfo {eid} {A45}
		(\bibinfo {year} {2011})},\ \Eprint {https://arxiv.org/abs/1012.5654}
	{arXiv:1012.5654 [astro-ph.SR]} \BibitemShut {NoStop}%
	\bibitem [{\citenamefont {{Huppenkothen}}\ \emph
		{et~al.}(2014{\natexlab{a}})\citenamefont {{Huppenkothen}}, \citenamefont
		{{Heil}}, \citenamefont {{Watts}},\ and\ \citenamefont
		{{G{\"o}{\u{g}}{\"u}{\textcommabelow s}}}}]{2014Huppen}%
	\BibitemOpen
	\bibfield  {author} {\bibinfo {author} {\bibfnamefont {D.}~\bibnamefont
			{{Huppenkothen}}}, \bibinfo {author} {\bibfnamefont {L.~M.}\ \bibnamefont
			{{Heil}}}, \bibinfo {author} {\bibfnamefont {A.~L.}\ \bibnamefont
			{{Watts}}},\ and\ \bibinfo {author} {\bibfnamefont {E.}~\bibnamefont
			{{G{\"o}{\u{g}}{\"u}{\textcommabelow s}}}},\ }\bibfield  {title} {\bibinfo
		{title} {{Quasi-periodic Oscillations in Short Recurring Bursts of Magnetars
				SGR 1806-20 and SGR 1900+14 Observed with RXTE}},\ }\href
	{https://doi.org/10.1088/0004-637X/795/2/114} {\bibfield  {journal} {\bibinfo
			{journal} {\apj}\ }\textbf {\bibinfo {volume} {795}},\ \bibinfo {eid} {114}
		(\bibinfo {year} {2014}{\natexlab{a}})},\ \Eprint
	{https://arxiv.org/abs/1409.7642} {arXiv:1409.7642 [astro-ph.HE]}
	\BibitemShut {NoStop}%
	\bibitem [{\citenamefont {{Huppenkothen}}\ \emph
		{et~al.}(2014{\natexlab{b}})\citenamefont {{Huppenkothen}}, \citenamefont
		{{D'Angelo}}, \citenamefont {{Watts}}, \citenamefont {{Heil}}, \citenamefont
		{{van der Klis}}, \citenamefont {{van der Horst}}, \citenamefont
		{{Kouveliotou}}, \citenamefont {{Baring}}, \citenamefont
		{{G{\"o}{\u{g}}{\"u}{\textcommabelow s}}}, \citenamefont {{Granot}},
		\citenamefont {{Kaneko}}, \citenamefont {{Lin}}, \citenamefont {{von
				Kienlin}},\ and\ \citenamefont {{Younes}}}]{2014Huppenkothen}%
	\BibitemOpen
	\bibfield  {author} {\bibinfo {author} {\bibfnamefont {D.}~\bibnamefont
			{{Huppenkothen}}}, \bibinfo {author} {\bibfnamefont {C.}~\bibnamefont
			{{D'Angelo}}}, \bibinfo {author} {\bibfnamefont {A.~L.}\ \bibnamefont
			{{Watts}}}, \bibinfo {author} {\bibfnamefont {L.}~\bibnamefont {{Heil}}},
		\bibinfo {author} {\bibfnamefont {M.}~\bibnamefont {{van der Klis}}},
		\bibinfo {author} {\bibfnamefont {A.~J.}\ \bibnamefont {{van der Horst}}},
		\bibinfo {author} {\bibfnamefont {C.}~\bibnamefont {{Kouveliotou}}}, \bibinfo
		{author} {\bibfnamefont {M.~G.}\ \bibnamefont {{Baring}}}, \bibinfo {author}
		{\bibfnamefont {E.}~\bibnamefont {{G{\"o}{\u{g}}{\"u}{\textcommabelow s}}}},
		\bibinfo {author} {\bibfnamefont {J.}~\bibnamefont {{Granot}}}, \bibinfo
		{author} {\bibfnamefont {Y.}~\bibnamefont {{Kaneko}}}, \bibinfo {author}
		{\bibfnamefont {L.}~\bibnamefont {{Lin}}}, \bibinfo {author} {\bibfnamefont
			{A.}~\bibnamefont {{von Kienlin}}},\ and\ \bibinfo {author} {\bibfnamefont
			{G.}~\bibnamefont {{Younes}}},\ }\bibfield  {title} {\bibinfo {title}
		{{Quasi-periodic Oscillations in Short Recurring Bursts of the Soft Gamma
				Repeater J1550-5418}},\ }\href {https://doi.org/10.1088/0004-637X/787/2/128}
	{\bibfield  {journal} {\bibinfo  {journal} {\apj}\ }\textbf {\bibinfo
			{volume} {787}},\ \bibinfo {eid} {128} (\bibinfo {year}
		{2014}{\natexlab{b}})},\ \Eprint {https://arxiv.org/abs/1404.2756}
	{arXiv:1404.2756 [astro-ph.HE]} \BibitemShut {NoStop}%
	\bibitem [{\citenamefont {{Pumpe}}\ \emph {et~al.}(2018)\citenamefont
		{{Pumpe}}, \citenamefont {{Gabler}}, \citenamefont {{Steininger}},\ and\
		\citenamefont {{En{\ss}lin}}}]{2018Pumpe}%
	\BibitemOpen
	\bibfield  {author} {\bibinfo {author} {\bibfnamefont {D.}~\bibnamefont
			{{Pumpe}}}, \bibinfo {author} {\bibfnamefont {M.}~\bibnamefont {{Gabler}}},
		\bibinfo {author} {\bibfnamefont {T.}~\bibnamefont {{Steininger}}},\ and\
		\bibinfo {author} {\bibfnamefont {T.~A.}\ \bibnamefont {{En{\ss}lin}}},\
	}\bibfield  {title} {\bibinfo {title} {{Search for quasi-periodic signals in
				magnetar giant flares. Bayesian inspection of SGR 1806-20 and SGR 1900+14}},\
	}\href {https://doi.org/10.1051/0004-6361/201731800} {\bibfield  {journal}
		{\bibinfo  {journal} {\aap}\ }\textbf {\bibinfo {volume} {610}},\ \bibinfo
		{eid} {A61} (\bibinfo {year} {2018})},\ \Eprint
	{https://arxiv.org/abs/1708.05702} {arXiv:1708.05702 [astro-ph.HE]}
	\BibitemShut {NoStop}%
	\bibitem [{\citenamefont {{Hansen}}\ and\ \citenamefont
		{{Cioffi}}(1980)}]{1980Hansen}%
	\BibitemOpen
	\bibfield  {author} {\bibinfo {author} {\bibfnamefont {C.~J.}\ \bibnamefont
			{{Hansen}}}\ and\ \bibinfo {author} {\bibfnamefont {D.~F.}\ \bibnamefont
			{{Cioffi}}},\ }\bibfield  {title} {\bibinfo {title} {{Torsional oscillations
				in neutron star crusts}},\ }\href {https://doi.org/10.1086/158031} {\bibfield
		{journal} {\bibinfo  {journal} {\apj}\ }\textbf {\bibinfo {volume} {238}},\
		\bibinfo {pages} {740} (\bibinfo {year} {1980})}\BibitemShut {NoStop}%
	\bibitem [{\citenamefont {{Schumaker}}\ and\ \citenamefont
		{{Thorne}}(1983)}]{1983ST}%
	\BibitemOpen
	\bibfield  {author} {\bibinfo {author} {\bibfnamefont {B.~L.}\ \bibnamefont
			{{Schumaker}}}\ and\ \bibinfo {author} {\bibfnamefont {K.~S.}\ \bibnamefont
			{{Thorne}}},\ }\bibfield  {title} {\bibinfo {title} {{Torsional oscillations
				of neutron stars}},\ }\href {https://doi.org/10.1093/mnras/203.2.457}
	{\bibfield  {journal} {\bibinfo  {journal} {\mnras}\ }\textbf {\bibinfo
			{volume} {203}},\ \bibinfo {pages} {457} (\bibinfo {year}
		{1983})}\BibitemShut {NoStop}%
	\bibitem [{\citenamefont {{McDermott}}\ \emph {et~al.}(1988)\citenamefont
		{{McDermott}}, \citenamefont {{van Horn}},\ and\ \citenamefont
		{{Hansen}}}]{1988McDermott}%
	\BibitemOpen
	\bibfield  {author} {\bibinfo {author} {\bibfnamefont {P.~N.}\ \bibnamefont
			{{McDermott}}}, \bibinfo {author} {\bibfnamefont {H.~M.}\ \bibnamefont {{van
					Horn}}},\ and\ \bibinfo {author} {\bibfnamefont {C.~J.}\ \bibnamefont
			{{Hansen}}},\ }\bibfield  {title} {\bibinfo {title} {{Nonradial Oscillations
				of Neutron Stars}},\ }\href {https://doi.org/10.1086/166044} {\bibfield
		{journal} {\bibinfo  {journal} {\apj}\ }\textbf {\bibinfo {volume} {325}},\
		\bibinfo {pages} {725} (\bibinfo {year} {1988})}\BibitemShut {NoStop}%
	\bibitem [{\citenamefont {{Samuelsson}}\ and\ \citenamefont
		{{Andersson}}(2007)}]{2007Samuel}%
	\BibitemOpen
	\bibfield  {author} {\bibinfo {author} {\bibfnamefont {L.}~\bibnamefont
			{{Samuelsson}}}\ and\ \bibinfo {author} {\bibfnamefont {N.}~\bibnamefont
			{{Andersson}}},\ }\bibfield  {title} {\bibinfo {title} {{Neutron star
				asteroseismology. Axial crust oscillations in the Cowling approximation}},\
	}\href {https://doi.org/10.1111/j.1365-2966.2006.11147.x} {\bibfield
		{journal} {\bibinfo  {journal} {\mnras}\ }\textbf {\bibinfo {volume} {374}},\
		\bibinfo {pages} {256} (\bibinfo {year} {2007})},\ \Eprint
	{https://arxiv.org/abs/astro-ph/0609265} {arXiv:astro-ph/0609265 [astro-ph]}
	\BibitemShut {NoStop}%
	\bibitem [{\citenamefont {{Andersson}}\ \emph {et~al.}(2009)\citenamefont
		{{Andersson}}, \citenamefont {{Glampedakis}},\ and\ \citenamefont
		{{Samuelsson}}}]{2009Andersson}%
	\BibitemOpen
	\bibfield  {author} {\bibinfo {author} {\bibfnamefont {N.}~\bibnamefont
			{{Andersson}}}, \bibinfo {author} {\bibfnamefont {K.}~\bibnamefont
			{{Glampedakis}}},\ and\ \bibinfo {author} {\bibfnamefont {L.}~\bibnamefont
			{{Samuelsson}}},\ }\bibfield  {title} {\bibinfo {title} {{Superfluid
				signatures in magnetar seismology}},\ }\href
	{https://doi.org/10.1111/j.1365-2966.2009.14734.x} {\bibfield  {journal}
		{\bibinfo  {journal} {\mnras}\ }\textbf {\bibinfo {volume} {396}},\ \bibinfo
		{pages} {894} (\bibinfo {year} {2009})},\ \Eprint
	{https://arxiv.org/abs/0812.2417} {arXiv:0812.2417 [astro-ph]} \BibitemShut
	{NoStop}%
	\bibitem [{\citenamefont {{Sotani}}\ \emph {et~al.}(2012)\citenamefont
		{{Sotani}}, \citenamefont {{Nakazato}}, \citenamefont {{Iida}},\ and\
		\citenamefont {{Oyamatsu}}}]{2012Sotani}%
	\BibitemOpen
	\bibfield  {author} {\bibinfo {author} {\bibfnamefont {H.}~\bibnamefont
			{{Sotani}}}, \bibinfo {author} {\bibfnamefont {K.}~\bibnamefont
			{{Nakazato}}}, \bibinfo {author} {\bibfnamefont {K.}~\bibnamefont {{Iida}}},\
		and\ \bibinfo {author} {\bibfnamefont {K.}~\bibnamefont {{Oyamatsu}}},\
	}\bibfield  {title} {\bibinfo {title} {{Probing the Equation of State of
				Nuclear Matter via Neutron Star Asteroseismology}},\ }\href
	{https://doi.org/10.1103/PhysRevLett.108.201101} {\bibfield  {journal}
		{\bibinfo  {journal} {\prl}\ }\textbf {\bibinfo {volume} {108}},\ \bibinfo
		{eid} {201101} (\bibinfo {year} {2012})},\ \Eprint
	{https://arxiv.org/abs/1202.6242} {arXiv:1202.6242 [astro-ph.HE]}
	\BibitemShut {NoStop}%
	\bibitem [{\citenamefont {{Sotani}}\ \emph
		{et~al.}(2013{\natexlab{a}})\citenamefont {{Sotani}}, \citenamefont
		{{Nakazato}}, \citenamefont {{Iida}},\ and\ \citenamefont
		{{Oyamatsu}}}]{2013aSotani}%
	\BibitemOpen
	\bibfield  {author} {\bibinfo {author} {\bibfnamefont {H.}~\bibnamefont
			{{Sotani}}}, \bibinfo {author} {\bibfnamefont {K.}~\bibnamefont
			{{Nakazato}}}, \bibinfo {author} {\bibfnamefont {K.}~\bibnamefont {{Iida}}},\
		and\ \bibinfo {author} {\bibfnamefont {K.}~\bibnamefont {{Oyamatsu}}},\
	}\bibfield  {title} {\bibinfo {title} {{Effect of superfluidity on neutron
				star oscillations}},\ }\href {https://doi.org/10.1093/mnrasl/sls006}
	{\bibfield  {journal} {\bibinfo  {journal} {\mnras}\ }\textbf {\bibinfo
			{volume} {428}},\ \bibinfo {pages} {L21} (\bibinfo {year}
		{2013}{\natexlab{a}})},\ \Eprint {https://arxiv.org/abs/1210.0955}
	{arXiv:1210.0955 [astro-ph.HE]} \BibitemShut {NoStop}%
	\bibitem [{\citenamefont {{Sotani}}\ \emph
		{et~al.}(2013{\natexlab{b}})\citenamefont {{Sotani}}, \citenamefont
		{{Nakazato}}, \citenamefont {{Iida}},\ and\ \citenamefont
		{{Oyamatsu}}}]{2013Sotani}%
	\BibitemOpen
	\bibfield  {author} {\bibinfo {author} {\bibfnamefont {H.}~\bibnamefont
			{{Sotani}}}, \bibinfo {author} {\bibfnamefont {K.}~\bibnamefont
			{{Nakazato}}}, \bibinfo {author} {\bibfnamefont {K.}~\bibnamefont {{Iida}}},\
		and\ \bibinfo {author} {\bibfnamefont {K.}~\bibnamefont {{Oyamatsu}}},\
	}\bibfield  {title} {\bibinfo {title} {{Possible constraints on the density
				dependence of the nuclear symmetry energy from quasi-periodic oscillations in
				soft gamma repeaters}},\ }\href {https://doi.org/10.1093/mnras/stt1152}
	{\bibfield  {journal} {\bibinfo  {journal} {\mnras}\ }\textbf {\bibinfo
			{volume} {434}},\ \bibinfo {pages} {2060} (\bibinfo {year}
		{2013}{\natexlab{b}})},\ \Eprint {https://arxiv.org/abs/1303.4500}
	{arXiv:1303.4500 [astro-ph.HE]} \BibitemShut {NoStop}%
	\bibitem [{\citenamefont {{Sotani}}(2016)}]{2016Sotani}%
	\BibitemOpen
	\bibfield  {author} {\bibinfo {author} {\bibfnamefont {H.}~\bibnamefont
			{{Sotani}}},\ }\bibfield  {title} {\bibinfo {title} {{Empirical formula of
				crustal torsional oscillations}},\ }\href
	{https://doi.org/10.1103/PhysRevD.93.044059} {\bibfield  {journal} {\bibinfo
			{journal} {\prd}\ }\textbf {\bibinfo {volume} {93}},\ \bibinfo {eid} {044059}
		(\bibinfo {year} {2016})},\ \Eprint {https://arxiv.org/abs/1602.04558}
	{arXiv:1602.04558 [astro-ph.HE]} \BibitemShut {NoStop}%
	\bibitem [{\citenamefont {{Sotani}}\ \emph
		{et~al.}(2017{\natexlab{a}})\citenamefont {{Sotani}}, \citenamefont
		{{Iida}},\ and\ \citenamefont {{Oyamatsu}}}]{2017aSotani}%
	\BibitemOpen
	\bibfield  {author} {\bibinfo {author} {\bibfnamefont {H.}~\bibnamefont
			{{Sotani}}}, \bibinfo {author} {\bibfnamefont {K.}~\bibnamefont {{Iida}}},\
		and\ \bibinfo {author} {\bibfnamefont {K.}~\bibnamefont {{Oyamatsu}}},\
	}\bibfield  {title} {\bibinfo {title} {{Probing nuclear bubble structure via
				neutron star asteroseismology}},\ }\href
	{https://doi.org/10.1093/mnras/stw2575} {\bibfield  {journal} {\bibinfo
			{journal} {\mnras}\ }\textbf {\bibinfo {volume} {464}},\ \bibinfo {pages}
		{3101} (\bibinfo {year} {2017}{\natexlab{a}})},\ \Eprint
	{https://arxiv.org/abs/1609.01802} {arXiv:1609.01802 [astro-ph.HE]}
	\BibitemShut {NoStop}%
	\bibitem [{\citenamefont {{Sotani}}\ \emph
		{et~al.}(2017{\natexlab{b}})\citenamefont {{Sotani}}, \citenamefont
		{{Iida}},\ and\ \citenamefont {{Oyamatsu}}}]{2017Sotani}%
	\BibitemOpen
	\bibfield  {author} {\bibinfo {author} {\bibfnamefont {H.}~\bibnamefont
			{{Sotani}}}, \bibinfo {author} {\bibfnamefont {K.}~\bibnamefont {{Iida}}},\
		and\ \bibinfo {author} {\bibfnamefont {K.}~\bibnamefont {{Oyamatsu}}},\
	}\bibfield  {title} {\bibinfo {title} {{Probing crustal structures from
				neutron star compactness}},\ }\href {https://doi.org/10.1093/mnras/stx1510}
	{\bibfield  {journal} {\bibinfo  {journal} {\mnras}\ }\textbf {\bibinfo
			{volume} {470}},\ \bibinfo {pages} {4397} (\bibinfo {year}
		{2017}{\natexlab{b}})},\ \Eprint {https://arxiv.org/abs/1706.04736}
	{arXiv:1706.04736 [astro-ph.HE]} \BibitemShut {NoStop}%
	\bibitem [{\citenamefont {{Sotani}}\ \emph {et~al.}(2018)\citenamefont
		{{Sotani}}, \citenamefont {{Iida}},\ and\ \citenamefont
		{{Oyamatsu}}}]{2018Sotani}%
	\BibitemOpen
	\bibfield  {author} {\bibinfo {author} {\bibfnamefont {H.}~\bibnamefont
			{{Sotani}}}, \bibinfo {author} {\bibfnamefont {K.}~\bibnamefont {{Iida}}},\
		and\ \bibinfo {author} {\bibfnamefont {K.}~\bibnamefont {{Oyamatsu}}},\
	}\bibfield  {title} {\bibinfo {title} {{Constraints on the nuclear equation
				of state and the neutron star structure from crustal torsional
				oscillations}},\ }\href {https://doi.org/10.1093/mnras/sty1755} {\bibfield
		{journal} {\bibinfo  {journal} {\mnras}\ }\textbf {\bibinfo {volume} {479}},\
		\bibinfo {pages} {4735} (\bibinfo {year} {2018})},\ \Eprint
	{https://arxiv.org/abs/1807.00528} {arXiv:1807.00528 [astro-ph.HE]}
	\BibitemShut {NoStop}%
	\bibitem [{\citenamefont {{Sotani}}\ \emph {et~al.}(2019)\citenamefont
		{{Sotani}}, \citenamefont {{Iida}},\ and\ \citenamefont
		{{Oyamatsu}}}]{2019Sotani}%
	\BibitemOpen
	\bibfield  {author} {\bibinfo {author} {\bibfnamefont {H.}~\bibnamefont
			{{Sotani}}}, \bibinfo {author} {\bibfnamefont {K.}~\bibnamefont {{Iida}}},\
		and\ \bibinfo {author} {\bibfnamefont {K.}~\bibnamefont {{Oyamatsu}}},\
	}\bibfield  {title} {\bibinfo {title} {{Astrophysical implications of
				double-layer torsional oscillations in a neutron star crust as a lasagna
				sandwich}},\ }\href {https://doi.org/10.1093/mnras/stz2385} {\bibfield
		{journal} {\bibinfo  {journal} {\mnras}\ }\textbf {\bibinfo {volume} {489}},\
		\bibinfo {pages} {3022} (\bibinfo {year} {2019})},\ \Eprint
	{https://arxiv.org/abs/1906.06999} {arXiv:1906.06999 [astro-ph.HE]}
	\BibitemShut {NoStop}%
	\bibitem [{\citenamefont {{Kozhberov}}\ and\ \citenamefont
		{{Yakovlev}}(2020)}]{2020KY}%
	\BibitemOpen
	\bibfield  {author} {\bibinfo {author} {\bibfnamefont {A.~A.}\ \bibnamefont
			{{Kozhberov}}}\ and\ \bibinfo {author} {\bibfnamefont {D.~G.}\ \bibnamefont
			{{Yakovlev}}},\ }\bibfield  {title} {\bibinfo {title} {{Deformed crystals and
				torsional oscillations of neutron star crust}},\ }\href
	{https://doi.org/10.1093/mnras/staa2715} {\bibfield  {journal} {\bibinfo
			{journal} {\mnras}\ }\textbf {\bibinfo {volume} {498}},\ \bibinfo {pages}
		{5149} (\bibinfo {year} {2020})},\ \Eprint {https://arxiv.org/abs/2009.04952}
	{arXiv:2009.04952 [astro-ph.HE]} \BibitemShut {NoStop}%
	\bibitem [{\citenamefont {{Yakovlev}}(2023{\natexlab{a}})}]{2023Yak2}%
	\BibitemOpen
	\bibfield  {author} {\bibinfo {author} {\bibfnamefont {D.~G.}\ \bibnamefont
			{{Yakovlev}}},\ }\bibfield  {title} {\bibinfo {title} {{Self-similarity
				relations for torsional oscillations of neutron stars}},\ }\href
	{https://doi.org/10.1093/mnras/stac2871} {\bibfield  {journal} {\bibinfo
			{journal} {\mnras}\ }\textbf {\bibinfo {volume} {518}},\ \bibinfo {pages}
		{1148} (\bibinfo {year} {2023}{\natexlab{a}})},\ \Eprint
	{https://arxiv.org/abs/2210.02931} {arXiv:2210.02931 [astro-ph.SR]}
	\BibitemShut {NoStop}%
	\bibitem [{\citenamefont {{Levin}}(2006)}]{2006Levin}%
	\BibitemOpen
	\bibfield  {author} {\bibinfo {author} {\bibfnamefont {Y.}~\bibnamefont
			{{Levin}}},\ }\bibfield  {title} {\bibinfo {title} {{QPOs during magnetar
				flares are not driven by mechanical normal modes of the crust}},\ }\href
	{https://doi.org/10.1111/j.1745-3933.2006.00155.x} {\bibfield  {journal}
		{\bibinfo  {journal} {\mnras}\ }\textbf {\bibinfo {volume} {368}},\ \bibinfo
		{pages} {L35} (\bibinfo {year} {2006})},\ \Eprint
	{https://arxiv.org/abs/astro-ph/0601020} {arXiv:astro-ph/0601020 [astro-ph]}
	\BibitemShut {NoStop}%
	\bibitem [{\citenamefont {{Glampedakis}}\ \emph {et~al.}(2006)\citenamefont
		{{Glampedakis}}, \citenamefont {{Samuelsson}},\ and\ \citenamefont
		{{Andersson}}}]{2006Glampeda}%
	\BibitemOpen
	\bibfield  {author} {\bibinfo {author} {\bibfnamefont {K.}~\bibnamefont
			{{Glampedakis}}}, \bibinfo {author} {\bibfnamefont {L.}~\bibnamefont
			{{Samuelsson}}},\ and\ \bibinfo {author} {\bibfnamefont {N.}~\bibnamefont
			{{Andersson}}},\ }\bibfield  {title} {\bibinfo {title} {{Elastic or magnetic?
				A toy model for global magnetar oscillations with implications for
				quasi-periodic oscillations during flares}},\ }\href
	{https://doi.org/10.1111/j.1745-3933.2006.00211.x} {\bibfield  {journal}
		{\bibinfo  {journal} {\mnras}\ }\textbf {\bibinfo {volume} {371}},\ \bibinfo
		{pages} {L74} (\bibinfo {year} {2006})},\ \Eprint
	{https://arxiv.org/abs/astro-ph/0605461} {arXiv:astro-ph/0605461 [astro-ph]}
	\BibitemShut {NoStop}%
	\bibitem [{\citenamefont {{Sotani}}\ \emph {et~al.}(2007)\citenamefont
		{{Sotani}}, \citenamefont {{Kokkotas}},\ and\ \citenamefont
		{{Stergioulas}}}]{2007Sotani}%
	\BibitemOpen
	\bibfield  {author} {\bibinfo {author} {\bibfnamefont {H.}~\bibnamefont
			{{Sotani}}}, \bibinfo {author} {\bibfnamefont {K.~D.}\ \bibnamefont
			{{Kokkotas}}},\ and\ \bibinfo {author} {\bibfnamefont {N.}~\bibnamefont
			{{Stergioulas}}},\ }\bibfield  {title} {\bibinfo {title} {{Torsional
				oscillations of relativistic stars with dipole magnetic fields}},\ }\href
	{https://doi.org/10.1111/j.1365-2966.2006.11304.x} {\bibfield  {journal}
		{\bibinfo  {journal} {\mnras}\ }\textbf {\bibinfo {volume} {375}},\ \bibinfo
		{pages} {261} (\bibinfo {year} {2007})},\ \Eprint
	{https://arxiv.org/abs/astro-ph/0608626} {arXiv:astro-ph/0608626 [astro-ph]}
	\BibitemShut {NoStop}%
	\bibitem [{\citenamefont {{Levin}}(2007)}]{2007Levin}%
	\BibitemOpen
	\bibfield  {author} {\bibinfo {author} {\bibfnamefont {Y.}~\bibnamefont
			{{Levin}}},\ }\bibfield  {title} {\bibinfo {title} {{On the theory of
				magnetar QPOs}},\ }\href {https://doi.org/10.1111/j.1365-2966.2007.11582.x}
	{\bibfield  {journal} {\bibinfo  {journal} {\mnras}\ }\textbf {\bibinfo
			{volume} {377}},\ \bibinfo {pages} {159} (\bibinfo {year} {2007})},\ \Eprint
	{https://arxiv.org/abs/astro-ph/0612725} {arXiv:astro-ph/0612725 [astro-ph]}
	\BibitemShut {NoStop}%
	\bibitem [{\citenamefont {{Sotani}}\ \emph {et~al.}(2008)\citenamefont
		{{Sotani}}, \citenamefont {{Kokkotas}},\ and\ \citenamefont
		{{Stergioulas}}}]{2008Sotani}%
	\BibitemOpen
	\bibfield  {author} {\bibinfo {author} {\bibfnamefont {H.}~\bibnamefont
			{{Sotani}}}, \bibinfo {author} {\bibfnamefont {K.~D.}\ \bibnamefont
			{{Kokkotas}}},\ and\ \bibinfo {author} {\bibfnamefont {N.}~\bibnamefont
			{{Stergioulas}}},\ }\bibfield  {title} {\bibinfo {title} {{Alfv\'en
				quasi-periodic oscillations in magnetars}},\ }\href
	{https://doi.org/10.1111/j.1745-3933.2007.00420.x} {\bibfield  {journal}
		{\bibinfo  {journal} {\mnras}\ }\textbf {\bibinfo {volume} {385}},\ \bibinfo
		{pages} {L5} (\bibinfo {year} {2008})},\ \Eprint
	{https://arxiv.org/abs/0710.1113} {arXiv:0710.1113 [astro-ph]} \BibitemShut
	{NoStop}%
	\bibitem [{\citenamefont {{Lee}}(2008)}]{2008Lee}%
	\BibitemOpen
	\bibfield  {author} {\bibinfo {author} {\bibfnamefont {U.}~\bibnamefont
			{{Lee}}},\ }\bibfield  {title} {\bibinfo {title} {{Axisymmetric toroidal
				modes of magnetized neutron stars}},\ }\href
	{https://doi.org/10.1111/j.1365-2966.2008.12965.x} {\bibfield  {journal}
		{\bibinfo  {journal} {\mnras}\ }\textbf {\bibinfo {volume} {385}},\ \bibinfo
		{pages} {2069} (\bibinfo {year} {2008})},\ \Eprint
	{https://arxiv.org/abs/0710.4986} {arXiv:0710.4986 [astro-ph]} \BibitemShut
	{NoStop}%
	\bibitem [{\citenamefont {{Colaiuda}}\ \emph {et~al.}(2009)\citenamefont
		{{Colaiuda}}, \citenamefont {{Beyer}},\ and\ \citenamefont
		{{Kokkotas}}}]{2009Colaiuda}%
	\BibitemOpen
	\bibfield  {author} {\bibinfo {author} {\bibfnamefont {A.}~\bibnamefont
			{{Colaiuda}}}, \bibinfo {author} {\bibfnamefont {H.}~\bibnamefont
			{{Beyer}}},\ and\ \bibinfo {author} {\bibfnamefont {K.~D.}\ \bibnamefont
			{{Kokkotas}}},\ }\bibfield  {title} {\bibinfo {title} {{On the quasi-periodic
				oscillations in magnetars}},\ }\href
	{https://doi.org/10.1111/j.1365-2966.2009.14878.x} {\bibfield  {journal}
		{\bibinfo  {journal} {\mnras}\ }\textbf {\bibinfo {volume} {396}},\ \bibinfo
		{pages} {1441} (\bibinfo {year} {2009})},\ \Eprint
	{https://arxiv.org/abs/0902.1401} {arXiv:0902.1401 [astro-ph.HE]}
	\BibitemShut {NoStop}%
	\bibitem [{\citenamefont {{Cerd{\'a}-Dur{\'a}n}}\ \emph
		{et~al.}(2009)\citenamefont {{Cerd{\'a}-Dur{\'a}n}}, \citenamefont
		{{Stergioulas}},\ and\ \citenamefont {{Font}}}]{2009CD}%
	\BibitemOpen
	\bibfield  {author} {\bibinfo {author} {\bibfnamefont {P.}~\bibnamefont
			{{Cerd{\'a}-Dur{\'a}n}}}, \bibinfo {author} {\bibfnamefont {N.}~\bibnamefont
			{{Stergioulas}}},\ and\ \bibinfo {author} {\bibfnamefont {J.~A.}\
			\bibnamefont {{Font}}},\ }\bibfield  {title} {\bibinfo {title} {{Alfv{\'e}n
				QPOs in magnetars in the anelastic approximation}},\ }\href
	{https://doi.org/10.1111/j.1365-2966.2009.15056.x} {\bibfield  {journal}
		{\bibinfo  {journal} {\mnras}\ }\textbf {\bibinfo {volume} {397}},\ \bibinfo
		{pages} {1607} (\bibinfo {year} {2009})},\ \Eprint
	{https://arxiv.org/abs/0902.1472} {arXiv:0902.1472 [astro-ph.HE]}
	\BibitemShut {NoStop}%
	\bibitem [{\citenamefont {{Gabler}}\ \emph {et~al.}(2011)\citenamefont
		{{Gabler}}, \citenamefont {{Cerd{\'a} Dur{\'a}n}}, \citenamefont {{Font}},
		\citenamefont {{M{\"u}ller}},\ and\ \citenamefont
		{{Stergioulas}}}]{2011Gabler}%
	\BibitemOpen
	\bibfield  {author} {\bibinfo {author} {\bibfnamefont {M.}~\bibnamefont
			{{Gabler}}}, \bibinfo {author} {\bibfnamefont {P.}~\bibnamefont {{Cerd{\'a}
					Dur{\'a}n}}}, \bibinfo {author} {\bibfnamefont {J.~A.}\ \bibnamefont
			{{Font}}}, \bibinfo {author} {\bibfnamefont {E.}~\bibnamefont
			{{M{\"u}ller}}},\ and\ \bibinfo {author} {\bibfnamefont {N.}~\bibnamefont
			{{Stergioulas}}},\ }\bibfield  {title} {\bibinfo {title} {{Magneto-elastic
				oscillations and the damping of crustal shear modes in magnetars}},\ }\href
	{https://doi.org/10.1111/j.1745-3933.2010.00974.x} {\bibfield  {journal}
		{\bibinfo  {journal} {\mnras}\ }\textbf {\bibinfo {volume} {410}},\ \bibinfo
		{pages} {L37} (\bibinfo {year} {2011})},\ \Eprint
	{https://arxiv.org/abs/1007.0856} {arXiv:1007.0856 [astro-ph.HE]}
	\BibitemShut {NoStop}%
	\bibitem [{\citenamefont {{van Hoven}}\ and\ \citenamefont
		{{Levin}}(2011)}]{2011vanHoven}%
	\BibitemOpen
	\bibfield  {author} {\bibinfo {author} {\bibfnamefont {M.}~\bibnamefont {{van
					Hoven}}}\ and\ \bibinfo {author} {\bibfnamefont {Y.}~\bibnamefont
			{{Levin}}},\ }\bibfield  {title} {\bibinfo {title} {{Magnetar oscillations -
				I. Strongly coupled dynamics of the crust and the core}},\ }\href
	{https://doi.org/10.1111/j.1365-2966.2010.17499.x} {\bibfield  {journal}
		{\bibinfo  {journal} {\mnras}\ }\textbf {\bibinfo {volume} {410}},\ \bibinfo
		{pages} {1036} (\bibinfo {year} {2011})},\ \Eprint
	{https://arxiv.org/abs/1006.0348} {arXiv:1006.0348 [astro-ph.HE]}
	\BibitemShut {NoStop}%
	\bibitem [{\citenamefont {{Colaiuda}}\ and\ \citenamefont
		{{Kokkotas}}(2011)}]{2011Colaiuda}%
	\BibitemOpen
	\bibfield  {author} {\bibinfo {author} {\bibfnamefont {A.}~\bibnamefont
			{{Colaiuda}}}\ and\ \bibinfo {author} {\bibfnamefont {K.~D.}\ \bibnamefont
			{{Kokkotas}}},\ }\bibfield  {title} {\bibinfo {title} {{Magnetar oscillations
				in the presence of a crust}},\ }\href
	{https://doi.org/10.1111/j.1365-2966.2011.18602.x} {\bibfield  {journal}
		{\bibinfo  {journal} {\mnras}\ }\textbf {\bibinfo {volume} {414}},\ \bibinfo
		{pages} {3014} (\bibinfo {year} {2011})},\ \Eprint
	{https://arxiv.org/abs/1012.3103} {arXiv:1012.3103 [gr-qc]} \BibitemShut
	{NoStop}%
	\bibitem [{\citenamefont {{van Hoven}}\ and\ \citenamefont
		{{Levin}}(2012)}]{2012vanHoven}%
	\BibitemOpen
	\bibfield  {author} {\bibinfo {author} {\bibfnamefont {M.}~\bibnamefont {{van
					Hoven}}}\ and\ \bibinfo {author} {\bibfnamefont {Y.}~\bibnamefont
			{{Levin}}},\ }\bibfield  {title} {\bibinfo {title} {{Magnetar oscillations -
				II. Spectral method}},\ }\href
	{https://doi.org/10.1111/j.1365-2966.2011.20177.x} {\bibfield  {journal}
		{\bibinfo  {journal} {\mnras}\ }\textbf {\bibinfo {volume} {420}},\ \bibinfo
		{pages} {3035} (\bibinfo {year} {2012})},\ \Eprint
	{https://arxiv.org/abs/1110.2107} {arXiv:1110.2107 [astro-ph.HE]}
	\BibitemShut {NoStop}%
	\bibitem [{\citenamefont {{Gabler}}\ \emph {et~al.}(2012)\citenamefont
		{{Gabler}}, \citenamefont {{Cerd{\'a}-Dur{\'a}n}}, \citenamefont
		{{Stergioulas}}, \citenamefont {{Font}},\ and\ \citenamefont
		{{M{\"u}ller}}}]{2012Gabler}%
	\BibitemOpen
	\bibfield  {author} {\bibinfo {author} {\bibfnamefont {M.}~\bibnamefont
			{{Gabler}}}, \bibinfo {author} {\bibfnamefont {P.}~\bibnamefont
			{{Cerd{\'a}-Dur{\'a}n}}}, \bibinfo {author} {\bibfnamefont {N.}~\bibnamefont
			{{Stergioulas}}}, \bibinfo {author} {\bibfnamefont {J.~A.}\ \bibnamefont
			{{Font}}},\ and\ \bibinfo {author} {\bibfnamefont {E.}~\bibnamefont
			{{M{\"u}ller}}},\ }\bibfield  {title} {\bibinfo {title} {{Magnetoelastic
				oscillations of neutron stars with dipolar magnetic fields}},\ }\href
	{https://doi.org/10.1111/j.1365-2966.2012.20454.x} {\bibfield  {journal}
		{\bibinfo  {journal} {\mnras}\ }\textbf {\bibinfo {volume} {421}},\ \bibinfo
		{pages} {2054} (\bibinfo {year} {2012})},\ \Eprint
	{https://arxiv.org/abs/1109.6233} {arXiv:1109.6233 [astro-ph.HE]}
	\BibitemShut {NoStop}%
	\bibitem [{\citenamefont {{Colaiuda}}\ and\ \citenamefont
		{{Kokkotas}}(2012)}]{2012Colaiuda}%
	\BibitemOpen
	\bibfield  {author} {\bibinfo {author} {\bibfnamefont {A.}~\bibnamefont
			{{Colaiuda}}}\ and\ \bibinfo {author} {\bibfnamefont {K.~D.}\ \bibnamefont
			{{Kokkotas}}},\ }\bibfield  {title} {\bibinfo {title} {{Coupled polar-axial
				magnetar oscillations}},\ }\href
	{https://doi.org/10.1111/j.1365-2966.2012.20919.x} {\bibfield  {journal}
		{\bibinfo  {journal} {\mnras}\ }\textbf {\bibinfo {volume} {423}},\ \bibinfo
		{pages} {811} (\bibinfo {year} {2012})},\ \Eprint
	{https://arxiv.org/abs/1112.3561} {arXiv:1112.3561 [astro-ph.HE]}
	\BibitemShut {NoStop}%
	\bibitem [{\citenamefont {{Gabler}}\ \emph
		{et~al.}(2013{\natexlab{a}})\citenamefont {{Gabler}}, \citenamefont
		{{Cerd{\'a}-Dur{\'a}n}}, \citenamefont {{Font}}, \citenamefont
		{{M{\"u}ller}},\ and\ \citenamefont {{Stergioulas}}}]{2013Gabler}%
	\BibitemOpen
	\bibfield  {author} {\bibinfo {author} {\bibfnamefont {M.}~\bibnamefont
			{{Gabler}}}, \bibinfo {author} {\bibfnamefont {P.}~\bibnamefont
			{{Cerd{\'a}-Dur{\'a}n}}}, \bibinfo {author} {\bibfnamefont {J.~A.}\
			\bibnamefont {{Font}}}, \bibinfo {author} {\bibfnamefont {E.}~\bibnamefont
			{{M{\"u}ller}}},\ and\ \bibinfo {author} {\bibfnamefont {N.}~\bibnamefont
			{{Stergioulas}}},\ }\bibfield  {title} {\bibinfo {title} {{Magneto-elastic
				oscillations of neutron stars: exploring different magnetic field
				configurations}},\ }\href {https://doi.org/10.1093/mnras/sts721} {\bibfield
		{journal} {\bibinfo  {journal} {\mnras}\ }\textbf {\bibinfo {volume} {430}},\
		\bibinfo {pages} {1811} (\bibinfo {year} {2013}{\natexlab{a}})},\ \Eprint
	{https://arxiv.org/abs/1208.6443} {arXiv:1208.6443 [astro-ph.SR]}
	\BibitemShut {NoStop}%
	\bibitem [{\citenamefont {{Gabler}}\ \emph
		{et~al.}(2013{\natexlab{b}})\citenamefont {{Gabler}}, \citenamefont
		{{Cerd{\'a}-Dur{\'a}n}}, \citenamefont {{Stergioulas}}, \citenamefont
		{{Font}},\ and\ \citenamefont {{M{\"u}ller}}}]{2013Gabler1}%
	\BibitemOpen
	\bibfield  {author} {\bibinfo {author} {\bibfnamefont {M.}~\bibnamefont
			{{Gabler}}}, \bibinfo {author} {\bibfnamefont {P.}~\bibnamefont
			{{Cerd{\'a}-Dur{\'a}n}}}, \bibinfo {author} {\bibfnamefont {N.}~\bibnamefont
			{{Stergioulas}}}, \bibinfo {author} {\bibfnamefont {J.~A.}\ \bibnamefont
			{{Font}}},\ and\ \bibinfo {author} {\bibfnamefont {E.}~\bibnamefont
			{{M{\"u}ller}}},\ }\bibfield  {title} {\bibinfo {title} {{Imprints of
				Superfluidity on Magnetoelastic Quasiperiodic Oscillations of Soft Gamma-Ray
				Repeaters}},\ }\href {https://doi.org/10.1103/PhysRevLett.111.211102}
	{\bibfield  {journal} {\bibinfo  {journal} {\prl}\ }\textbf {\bibinfo
			{volume} {111}},\ \bibinfo {eid} {211102} (\bibinfo {year}
		{2013}{\natexlab{b}})},\ \Eprint {https://arxiv.org/abs/1304.3566}
	{arXiv:1304.3566 [astro-ph.HE]} \BibitemShut {NoStop}%
	\bibitem [{\citenamefont {{Passamonti}}\ and\ \citenamefont
		{{Lander}}(2014)}]{2014Passamon}%
	\BibitemOpen
	\bibfield  {author} {\bibinfo {author} {\bibfnamefont {A.}~\bibnamefont
			{{Passamonti}}}\ and\ \bibinfo {author} {\bibfnamefont {S.~K.}\ \bibnamefont
			{{Lander}}},\ }\bibfield  {title} {\bibinfo {title} {{Quasi-periodic
				oscillations in superfluid magnetars}},\ }\href
	{https://doi.org/10.1093/mnras/stt2134} {\bibfield  {journal} {\bibinfo
			{journal} {\mnras}\ }\textbf {\bibinfo {volume} {438}},\ \bibinfo {pages}
		{156} (\bibinfo {year} {2014})},\ \Eprint {https://arxiv.org/abs/1307.3210}
	{arXiv:1307.3210 [astro-ph.SR]} \BibitemShut {NoStop}%
	\bibitem [{\citenamefont {{Link}}\ and\ \citenamefont {{van
				Eysden}}(2016)}]{2016Link}%
	\BibitemOpen
	\bibfield  {author} {\bibinfo {author} {\bibfnamefont {B.}~\bibnamefont
			{{Link}}}\ and\ \bibinfo {author} {\bibfnamefont {C.~A.}\ \bibnamefont {{van
					Eysden}}},\ }\bibfield  {title} {\bibinfo {title} {{Torsional Oscillations of
				a Magnetar with a Tangled Magnetic Field}},\ }\href
	{https://doi.org/10.3847/2041-8205/823/1/L1} {\bibfield  {journal} {\bibinfo
			{journal} {\apjl}\ }\textbf {\bibinfo {volume} {823}},\ \bibinfo {eid} {L1}
		(\bibinfo {year} {2016})},\ \Eprint {https://arxiv.org/abs/1604.02372}
	{arXiv:1604.02372 [astro-ph.HE]} \BibitemShut {NoStop}%
	\bibitem [{\citenamefont {{Gabler}}\ \emph {et~al.}(2016)\citenamefont
		{{Gabler}}, \citenamefont {{Cerd{\'a}-Dur{\'a}n}}, \citenamefont
		{{Stergioulas}}, \citenamefont {{Font}},\ and\ \citenamefont
		{{M{\"u}ller}}}]{2016Gabler}%
	\BibitemOpen
	\bibfield  {author} {\bibinfo {author} {\bibfnamefont {M.}~\bibnamefont
			{{Gabler}}}, \bibinfo {author} {\bibfnamefont {P.}~\bibnamefont
			{{Cerd{\'a}-Dur{\'a}n}}}, \bibinfo {author} {\bibfnamefont {N.}~\bibnamefont
			{{Stergioulas}}}, \bibinfo {author} {\bibfnamefont {J.~A.}\ \bibnamefont
			{{Font}}},\ and\ \bibinfo {author} {\bibfnamefont {E.}~\bibnamefont
			{{M{\"u}ller}}},\ }\bibfield  {title} {\bibinfo {title} {{Coherent
				magneto-elastic oscillations in superfluid magnetars}},\ }\href
	{https://doi.org/10.1093/mnras/stw1272} {\bibfield  {journal} {\bibinfo
			{journal} {\mnras}\ }\textbf {\bibinfo {volume} {460}},\ \bibinfo {pages}
		{4242} (\bibinfo {year} {2016})},\ \Eprint {https://arxiv.org/abs/1605.07638}
	{arXiv:1605.07638 [astro-ph.HE]} \BibitemShut {NoStop}%
	\bibitem [{\citenamefont {{Gabler}}\ \emph {et~al.}(2018)\citenamefont
		{{Gabler}}, \citenamefont {{Cerd{\'a}-Dur{\'a}n}}, \citenamefont
		{{Stergioulas}}, \citenamefont {{Font}},\ and\ \citenamefont
		{{M{\"u}ller}}}]{2018Gabler}%
	\BibitemOpen
	\bibfield  {author} {\bibinfo {author} {\bibfnamefont {M.}~\bibnamefont
			{{Gabler}}}, \bibinfo {author} {\bibfnamefont {P.}~\bibnamefont
			{{Cerd{\'a}-Dur{\'a}n}}}, \bibinfo {author} {\bibfnamefont {N.}~\bibnamefont
			{{Stergioulas}}}, \bibinfo {author} {\bibfnamefont {J.~A.}\ \bibnamefont
			{{Font}}},\ and\ \bibinfo {author} {\bibfnamefont {E.}~\bibnamefont
			{{M{\"u}ller}}},\ }\bibfield  {title} {\bibinfo {title} {{Constraining
				properties of high-density matter in neutron stars with magneto-elastic
				oscillations}},\ }\href {https://doi.org/10.1093/mnras/sty445} {\bibfield
		{journal} {\bibinfo  {journal} {\mnras}\ }\textbf {\bibinfo {volume} {476}},\
		\bibinfo {pages} {4199} (\bibinfo {year} {2018})},\ \Eprint
	{https://arxiv.org/abs/1710.02334} {arXiv:1710.02334 [astro-ph.HE]}
	\BibitemShut {NoStop}%
	\bibitem [{\citenamefont {{Cheung}}\ \emph {et~al.}(2024)\citenamefont
		{{Cheung}}, \citenamefont {{Lin}},\ and\ \citenamefont
		{{Chamel}}}]{2024Chamel}%
	\BibitemOpen
	\bibfield  {author} {\bibinfo {author} {\bibfnamefont {L.}~\bibnamefont
			{{Cheung}}}, \bibinfo {author} {\bibfnamefont {L.-M.}\ \bibnamefont
			{{Lin}}},\ and\ \bibinfo {author} {\bibfnamefont {N.}~\bibnamefont
			{{Chamel}}},\ }\bibfield  {title} {\bibinfo {title} {{Torsional oscillations
				of magnetized neutron stars: Impacts of Landau-Rabi quantization of electron
				motion}},\ }\href {https://doi.org/10.1103/PhysRevD.110.083021} {\bibfield
		{journal} {\bibinfo  {journal} {\prd}\ }\textbf {\bibinfo {volume} {110}},\
		\bibinfo {eid} {083021} (\bibinfo {year} {2024})},\ \Eprint
	{https://arxiv.org/abs/2312.05676} {arXiv:2312.05676 [astro-ph.HE]}
	\BibitemShut {NoStop}%
	\bibitem [{\citenamefont {{Shaisultanov}}\ and\ \citenamefont
		{{Eichler}}(2009)}]{2009SE}%
	\BibitemOpen
	\bibfield  {author} {\bibinfo {author} {\bibfnamefont {R.}~\bibnamefont
			{{Shaisultanov}}}\ and\ \bibinfo {author} {\bibfnamefont {D.}~\bibnamefont
			{{Eichler}}},\ }\bibfield  {title} {\bibinfo {title} {{What Magnetar
				Seismology Can Teach Us About Magnetic Fields}},\ }\href
	{https://doi.org/10.1088/0004-637X/702/1/L23} {\bibfield  {journal} {\bibinfo
			{journal} {\apjl}\ }\textbf {\bibinfo {volume} {702}},\ \bibinfo {pages}
		{L23} (\bibinfo {year} {2009})},\ \Eprint {https://arxiv.org/abs/0903.3319}
	{arXiv:0903.3319 [astro-ph.HE]} \BibitemShut {NoStop}%
	\bibitem [{\citenamefont {{Yakovlev}}(2023{\natexlab{b}})}]{2023Yak1}%
	\BibitemOpen
	\bibfield  {author} {\bibinfo {author} {\bibfnamefont {D.}~\bibnamefont
			{{Yakovlev}}},\ }\bibfield  {title} {\bibinfo {title} {{Zeeman Splitting of
				Torsional Oscillation Frequencies of Magnetars}},\ }\href
	{https://doi.org/10.3390/universe9120504} {\bibfield  {journal} {\bibinfo
			{journal} {Universe}\ }\textbf {\bibinfo {volume} {9}},\ \bibinfo {eid} {504}
		(\bibinfo {year} {2023}{\natexlab{b}})},\ \Eprint
	{https://arxiv.org/abs/2312.10022} {arXiv:2312.10022 [astro-ph.HE]}
	\BibitemShut {NoStop}%
	\bibitem [{\citenamefont {{Yakovlev}}(2024)}]{2024Yak}%
	\BibitemOpen
	\bibfield  {author} {\bibinfo {author} {\bibfnamefont {D.~G.}\ \bibnamefont
			{{Yakovlev}}},\ }\bibfield  {title} {\bibinfo {title} {{Powerful flares and
				magneto-elastic oscillations of magnetars}},\ }\href
	{https://doi.org/10.31857/S0044451024070125} {\bibfield  {journal} {\bibinfo
			{journal} {Zh. Exp. Teor. Fiz.}\ }\textbf {\bibinfo {volume} {166}},\
		\bibinfo {pages} {121} (\bibinfo {year} {2024})},\ \Eprint
	{https://arxiv.org/abs/2409.11178} {arXiv:2409.11178 [astro-ph]} \BibitemShut
	{NoStop}%
	\bibitem [{\citenamefont {{Arfken}}(1966)}]{1966Arfken}%
	\BibitemOpen
	\bibfield  {author} {\bibinfo {author} {\bibfnamefont {G.}~\bibnamefont
			{{Arfken}}},\ }\href@noop {} {\emph {\bibinfo {title} {{Mathematical Methods
					for Physicists}}}}\ (\bibinfo  {publisher} {Academic Press, New York},\
	\bibinfo {year} {1966})\BibitemShut {NoStop}%
	\bibitem [{\citenamefont {{Landau}}\ and\ \citenamefont
		{{Lifshitz}}(1976)}]{LL76}%
	\BibitemOpen
	\bibfield  {author} {\bibinfo {author} {\bibfnamefont {L.~D.}\ \bibnamefont
			{{Landau}}}\ and\ \bibinfo {author} {\bibfnamefont {E.~M.}\ \bibnamefont
			{{Lifshitz}}},\ }\href@noop {} {\emph {\bibinfo {title} {{Quantum
					Mechanics}}}}\ (\bibinfo  {publisher} {Oxford: Pergamon Press},\ \bibinfo
	{year} {1976})\BibitemShut {NoStop}%
	\bibitem [{\citenamefont {{Aguilera}}\ \emph {et~al.}(2008)\citenamefont
		{{Aguilera}}, \citenamefont {{Pons}},\ and\ \citenamefont
		{{Miralles}}}]{2008Aguilera}%
	\BibitemOpen
	\bibfield  {author} {\bibinfo {author} {\bibfnamefont {D.~N.}\ \bibnamefont
			{{Aguilera}}}, \bibinfo {author} {\bibfnamefont {J.~A.}\ \bibnamefont
			{{Pons}}},\ and\ \bibinfo {author} {\bibfnamefont {J.~A.}\ \bibnamefont
			{{Miralles}}},\ }\bibfield  {title} {\bibinfo {title} {{2D Cooling of
				magnetized neutron stars}},\ }\href
	{https://doi.org/10.1051/0004-6361:20078786} {\bibfield  {journal} {\bibinfo
			{journal} {\aap}\ }\textbf {\bibinfo {volume} {486}},\ \bibinfo {pages} {255}
		(\bibinfo {year} {2008})},\ \Eprint {https://arxiv.org/abs/0710.0854}
	{arXiv:0710.0854 [astro-ph]} \BibitemShut {NoStop}%
	\bibitem [{\citenamefont {{Potekhin}}\ \emph {et~al.}(2013)\citenamefont
		{{Potekhin}}, \citenamefont {{Fantina}}, \citenamefont {{Chamel}},
		\citenamefont {{Pearson}},\ and\ \citenamefont {{Goriely}}}]{BSk2013}%
	\BibitemOpen
	\bibfield  {author} {\bibinfo {author} {\bibfnamefont {A.~Y.}\ \bibnamefont
			{{Potekhin}}}, \bibinfo {author} {\bibfnamefont {A.~F.}\ \bibnamefont
			{{Fantina}}}, \bibinfo {author} {\bibfnamefont {N.}~\bibnamefont {{Chamel}}},
		\bibinfo {author} {\bibfnamefont {J.~M.}\ \bibnamefont {{Pearson}}},\ and\
		\bibinfo {author} {\bibfnamefont {S.}~\bibnamefont {{Goriely}}},\ }\bibfield
	{title} {\bibinfo {title} {{Analytical representations of unified equations
				of state for neutron-star matter}},\ }\href
	{https://doi.org/10.1051/0004-6361/201321697} {\bibfield  {journal} {\bibinfo
			{journal} {Astron. Astrophys.}\ }\textbf {\bibinfo {volume} {560}},\
		\bibinfo {eid} {A48} (\bibinfo {year} {2013})},\ \Eprint
	{https://arxiv.org/abs/1310.0049} {arXiv:1310.0049 [astro-ph.SR]}
	\BibitemShut {NoStop}%
	\bibitem [{\citenamefont {{Ogata}}\ and\ \citenamefont
		{{Ichimaru}}(1990)}]{1990Ogata}%
	\BibitemOpen
	\bibfield  {author} {\bibinfo {author} {\bibfnamefont {S.}~\bibnamefont
			{{Ogata}}}\ and\ \bibinfo {author} {\bibfnamefont {S.}~\bibnamefont
			{{Ichimaru}}},\ }\bibfield  {title} {\bibinfo {title} {{First-principles
				calculations of shear moduli for Monte Carlo-simulated Coulomb solids}},\
	}\href {https://doi.org/10.1103/PhysRevA.42.4867} {\bibfield  {journal}
		{\bibinfo  {journal} {\pra}\ }\textbf {\bibinfo {volume} {42}},\ \bibinfo
		{pages} {4867} (\bibinfo {year} {1990})}\BibitemShut {NoStop}%
	\bibitem [{\citenamefont {{Landau}}(1932)}]{1932Landau}%
	\BibitemOpen
	\bibfield  {author} {\bibinfo {author} {\bibfnamefont {L.~D.}\ \bibnamefont
			{{Landau}}},\ }\bibfield  {title} {\bibinfo {title} {{To the Stars theory}},\
	}\href@noop {} {\bibfield  {journal} {\bibinfo  {journal} {Phys. Zs. Sowjet}\
		}\textbf {\bibinfo {volume} {1}},\ \bibinfo {pages} {285} (\bibinfo {year}
		{1932})}\BibitemShut {NoStop}%
	\bibitem [{\citenamefont {{Yakovlev}}\ \emph {et~al.}(2013)\citenamefont
		{{Yakovlev}}, \citenamefont {{Haensel}}, \citenamefont {{Baym}},\ and\
		\citenamefont {{Pethick}}}]{2013Yak}%
	\BibitemOpen
	\bibfield  {author} {\bibinfo {author} {\bibfnamefont {D.~G.}\ \bibnamefont
			{{Yakovlev}}}, \bibinfo {author} {\bibfnamefont {P.}~\bibnamefont
			{{Haensel}}}, \bibinfo {author} {\bibfnamefont {G.}~\bibnamefont {{Baym}}},\
		and\ \bibinfo {author} {\bibfnamefont {C.}~\bibnamefont {{Pethick}}},\
	}\bibfield  {title} {\bibinfo {title} {{Lev Landau and the concept of neutron
				stars}},\ }\href {https://doi.org/10.3367/UFNe.0183.201303f.0307} {\bibfield
		{journal} {\bibinfo  {journal} {Physics Uspekhi}\ }\textbf {\bibinfo {volume}
			{56}},\ \bibinfo {eid} {289-295} (\bibinfo {year} {2013})},\ \Eprint
	{https://arxiv.org/abs/1210.0682} {arXiv:1210.0682 [physics.hist-ph]}
	\BibitemShut {NoStop}%
\end{thebibliography}

%

\end{document}